\documentclass[]{spie}  % default (US-letter). Use [a4paper] for A4.

\usepackage[english]{babel}
\usepackage{amsmath,amssymb}
\usepackage{graphicx}
\usepackage[percent]{overpic}
\graphicspath{{figs/}}
\usepackage{xcolor}
\usepackage[colorlinks=true, allcolors=blue, breaklinks=true]{hyperref}

\title{Upgrading LBTI/NOMIC with a quadruple annular groove phase mask and
GeoSnap detector for imaging nearby, habitable-zone exoplanets}

\author{%
Kevin Wagner\supit{a}, Manny Montoya\supit{a}, Steve Ertel\supit{a,b},
Jarron Leisenring\supit{a}, Pontus Forsberg\supit{c}, Samuel Ronayette\supit{d},
Andre Wong\supit{a}, Mikael Karlsson\supit{c}, Olivier Absil\supit{e},
Denis Defr\`ere\supit{f}, Markus Kasper\supit{g}, Jordan Stone\supit{h},
D\'aniel Apai\supit{a,i}, Laird Close\supit{a}, Jamie Dietrich\supit{j},
Ewan Douglas\supit{a}, Jamie Drew\supit{k}, Olivier Durney\supit{a}, Marina Fetisova\supit{l},
Kyran Grattan\supit{k}, Olivier Guyon\supit{a,m}, Jacob Isbell\supit{a},
Sebasti\'an Jorquera\supit{a}, Petri Karvinen\supit{l}, Markku Kuittinen\supit{l},
Herv\'e Le Coroller\supit{n}, Jared Males\supit{a}, Brittany Miles\supit{a},
Dillon O'Reilly\supit{k}, Eric Pantin\supit{d}, Sascha P.\ Quanz\supit{o,p},
Eckhart Spalding\supit{o,q}, Vivek Vijayakumar\supit{a}, Zach Werber\supit{a,r},
and S.\ Pete Worden\supit{a,k}
\skiplinehalf
\footnotesize
\supit{a}Steward Observatory, University of Arizona, 933 N.\ Cherry Ave., Tucson, AZ 85721, USA\\
\supit{b}Large Binocular Telescope Observatory, University of Arizona, Tucson, AZ 85721, USA\\
\supit{c}Uppsala University, Uppsala, Sweden\\
\supit{d}Universit\'e Paris-Saclay, CEA, Gif-sur-Yvette, France\\
\supit{e}STAR Institute, Universit\'e de Li\`ege, all\'ee du Six Ao\^ut 19c, 4000 Li\`ege, Belgium\\
\supit{f}KU Leuven, Leuven, Belgium\\
\supit{g}European Southern Observatory, Garching, Germany\\
\supit{h}University of Wyoming, Laramie, WY, USA\\
\supit{i}Lunar and Planetary Laboratory, University of Arizona, Tucson, AZ 85721, USA\\
\supit{j}Arizona State University, Tempe, AZ, USA\\
\supit{k}Breakthrough Initiatives, Steward Observatory, University of Arizona, 933 N.\ Cherry Ave., Tucson, AZ 85721, USA\\
\supit{l}University of Eastern Finland, Joensuu, Finland\\
\supit{m}National Astronomical Observatory of Japan, Mitaka, Tokyo, Japan\\
\supit{n}Laboratoire d'Astrophysique de Marseille, Marseille, France\\
\supit{o}ETH Z\"urich, Institute for Particle Physics and Astrophysics, Wolfgang-Pauli-Strasse~27, 8093 Z\"urich, Switzerland\\
\supit{p}ETH Z\"urich, Department of Earth and Planetary Sciences, Sonneggstrasse~5, 8092 Z\"urich, Switzerland\\
\supit{q}University of Sydney, Sydney, Australia\\
\supit{r}University of Hawai`i, Honolulu, HI, USA
}

\authorinfo{Send correspondence to K.W.: E-mail: kevinwagner@arizona.edu}

\begin{document}
\maketitle
\pagestyle{plain}\thispagestyle{plain}

\begin{abstract}
The Large Binocular Telescope Interferometer (LBTI)'s Nulling-Optimized Mid-Infrared Camera (NOMIC) is among the most capable thermal-infrared imaging systems available for high-contrast, high-angular-resolution astronomical observations. Here we describe two in-progress upgrades to LBTI/NOMIC: (1) the design, fabrication, and installation of a quadruple annular groove phase mask (Q-AGPM) coronagraph, and (2) the installation of a $13\,\mu$m-cutoff Teledyne GeoSnap array. The Q-AGPM is the first coronagraph to be installed within NOMIC and one of the first optimized for $N$-band ($\sim$11\,$\mu$m) observations. It places four annular groove phase masks on a single diamond substrate so that, in the LBTI dual-aperture imaging mode, each of the two telescope beams can be chopped between a pair of masks without loss of observing efficiency. The GeoSnap array will replace NOMIC's original AQUARIUS array, delivering higher quantum efficiency, larger well depth, faster and more linear readout, and freedom from the excess low-frequency noise that requires aggressive chopping. Together these upgrades substantially improve the achievable contrast and sensitivity at small angular separations. We also present a high-contrast Fizeau imaging sequence obtained with LBTI's new FFTCam fringe tracker, which confirms the interferometric gain over a single aperture through injection/recovery tests: relative to an equal-time single aperture exposure, the S/N\,=\,3 contrast is a factor of $\sim$2--4 deeper across $0.2$--$1''$, spanning the contrast- and background-limited regimes. Finally, we describe the role of the upgraded LBTI/NOMIC instrument within the Breakthrough Watch program at the University of Arizona, which aims to perform the deepest observations yet of the habitable zones of the nearest single Sun-like stars.
\end{abstract}


%%%%%%%%%%%%%%%%%%%%%%%%%%%%%%%%%%%%%%%%%%%%%%%%%%%%%%%%%%%%%%%%%%%%%%%%%%%%%%%%%
\section{INTRODUCTION}
\label{sec:intro}

\begin{figure}[t]
  \centering
  \includegraphics[width=\linewidth]{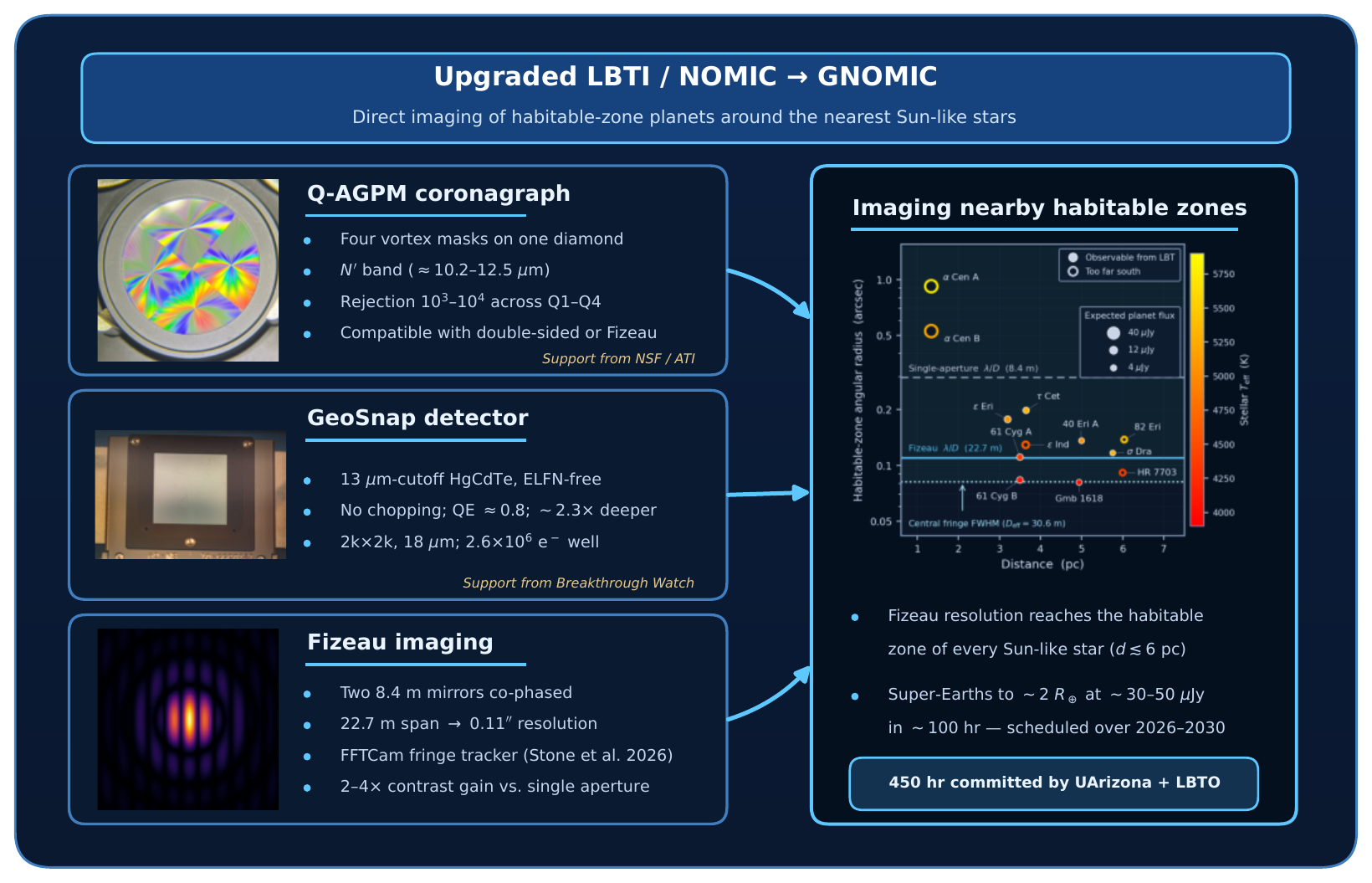}
  \caption[overview]{\label{fig:overview}
Overview of GNOMIC and its driving science goal. Three developments --- the quadruple annular groove phase mask (Q-AGPM) coronagraph, the $13\,\mu$m-cutoff GeoSnap HgCdTe detector, and co-phased Fizeau imaging with the two 8.4\,m LBT apertures --- combine to reach the small angular separations and deep contrasts required to directly image habitable-zone planets around the nearest single Sun-like stars. Representative performance figures for each component are annotated. The science panel plots the Earth-equivalent ($S_{\rm eff}\!=\!1$) habitable-zone angular radius --- the ``center'' column of Table~\ref{tab:targets}, $a_{\rm HZ}\!\approx\!\sqrt{L/L_\odot}$\,au --- versus distance for the target sample; symbol color encodes stellar effective temperature, symbol size the expected thermal flux of a temperate $\sim$$2\,R_\oplus$ planet, and filled versus open symbols denote stars that are observable from the LBT versus too far south (declination $\lesssim-30^\circ$).}
\end{figure}

Directly imaging temperate, terrestrial-mass planets around the nearest Sun-like stars is one of the defining goals of modern exoplanet science.\cite{astro2020, exostrategy2018} Over the past two decades, direct imaging has revealed and spectroscopically characterized young, self-luminous giant planets and protoplanets at wide separations from their host stars,\cite{chauvin2004, marois2008, macintosh2015, haffert2019} but the older, temperate, terrestrial-mass planets in the habitable zones of the nearest stars remain beyond the reach of current contrast limits. The contrast between such a planet and its host star is far more favorable in the thermal infrared than in reflected visible light: an Earth-sized $\sim$300\,K planet radiates most strongly near $10\,\mu$m, where the planet-to-star contrast is of order $10^{-7}$, compared with $\sim$$10^{-10}$ for reflected starlight at optical and near-infrared wavelengths.\cite{quanz2015, kasper2017, wagner2021} Ground-based $N$-band ($8$--$13\,\mu$m) imaging on large apertures is therefore a near-term route to detecting habitable-zone companions around the very closest stars, complementing the reflected-light approaches planned for future space missions and laying groundwork for upcoming mid-infrared facilities such as ELT/METIS and the proposed LIFE interferometer.\cite{brandl2021, quanz2022} Figure~\ref{fig:overview} summarizes the three enabling developments to the Large Binocular Telescope Interferometer (LBTI) Nulling-Optimized Mid-Infrared Camera (NOMIC) instrument\cite{hoffmann2014, hinz2016, ertel2020spie} described here and their shared science goal. We hereafter plan to refer to the upgraded instrument as GNOMIC, reflecting the primary detector upgrade to a $13\,\mu$m-cutoff Teledyne GeoSnap array and preserving the heritage and recognizability of NOMIC.

The potential for exoplanet imaging in the mid-IR was recently demonstrated by the Breakthrough Watch / New Earths in the Alpha Centauri Region (NEAR) experiment, which installed the first $\sim$11\,$\mu$m exoplanet-imaging system on the 8.2\,m Very Large Telescope and reached sub-Neptune sensitivity in roughly 100 hours of observations of the $\alpha$\,Centauri binary.\cite{kasper2017, wagner2021} Building on NEAR, our team has been working to make these capabilities routinely available on the Large Binocular Telescope (LBT) --- a $\sim$23\,m edge-to-edge facility built with a strong emphasis on mid-infrared performance, and the only existing prototype of a $30$\,m-class extremely large telescope.\cite{hinz2016, ertel2020} Operating the two 8.4\,m LBT apertures together in Fizeau imaging mode\cite{spalding, isbell2025, labeyrie1996, monnier2003} improves the angular resolution by roughly a factor of three relative to a single aperture. Combined with a state-of-the-art detector, the upgraded instrument is expected to gain roughly a factor of six in sensitivity over NEAR,\cite{ertel2022} sufficient to probe the habitable zones of the nearest single Sun-like stars such as $\epsilon$\,Eridani and $\tau$\,Ceti.\cite{wagner2021spie, ertel2022}

In this paper we describe two upgrades that together unlock this performance. Section~\ref{sec:opportunity} sets out the scientific opportunity $-$ resolved mid-infrared imaging of the habitable zones of the nearest Sun-like stars $-$ that motivates them. Section~\ref{sec:nomic} reviews the LBTI/NOMIC instrument and its imaging modes. Section~\ref{sec:qagpm} describes the quadruple annular groove phase mask (Q-AGPM), the first coronagraph to be installed inside NOMIC.\cite{forsberg2026} Section~\ref{sec:geosnap} describes replacement of the legacy AQUARIUS array with a $13\,\mu$m-cutoff Teledyne GeoSnap detector. Section~\ref{sec:fizeau} presents the Fizeau imaging mode and on-sky results obtained with the new FFTCam fringe tracker. Section~\ref{sec:breakthrough} places these upgrades and planned observations in the context of the Breakthrough Watch program, and Section~\ref{sec:summary} summarizes the status and expected science gains.

%%%%%%%%%%%%%%%%%%%%%%%%%%%%%%%%%%%%%%%%%%%%%%%%%%%%%%%%%%%%%%%%%%%%%%%%%%%%%%%%%
\section{THE MID-INFRARED OPPORTUNITY: RESOLVING NEARBY HABITABLE ZONES}
\label{sec:opportunity}

\subsection{Nearby planets in the thermal infrared}
A temperate, Earth-sized planet is intrinsically faint, but it is far less faint relative to its host star in the thermal infrared than in reflected light. A $\sim$300\,K planet emits most of its energy near $10\,\mu$m, and at these wavelengths the planet-to-star flux ratio for an Earth analog around a Sun-like star is of order $10^{-7}$, roughly three orders of magnitude more favorable than the $\sim$$10^{-10}$ contrast of reflected starlight in the visible and near-infrared.\cite{quanz2015} This relaxed contrast brings the direct detection of habitable-zone planets within reach of ground-based telescopes equipped with extreme adaptive optics and coronagraphy.\cite{macintosh2014, hinkley2011} The mid-infrared also carries unique diagnostic value: thermal emission constrains a planet's effective temperature and radius and provides access to key molecular bands. Within NOMIC's $N$-band window ($\sim$8--13\,$\mu$m), the accessible diagnostic is the $9.6\,\mu$m ozone band; the strong CO$_2$ band at $15\,\mu$m and other longer-wavelength features fall beyond it.\cite{desmarais2002, kaltenegger2017, schwieterman2018} This makes the mid-infrared highly complementary both to the reflected-light measurements envisioned for future flagship imaging missions\cite{cash2006, mamajek2024} and to proposed mid-infrared space interferometers such as LIFE.\cite{quanz2015, kopparapu2013, quanz2022, bracewell1978, leger1996}

\subsection{The angular-resolution challenge}
Ground-based $N$-band imaging faces two principal challenges. The first is the bright, fluctuating thermal background, which drives the need for tens of hours of integration per target, making independent second-epoch confirmation demanding. This is handled with well-established techniques such as rapid chopping and nodding and, going forward, the detector upgrade of Sec.~\ref{sec:geosnap}. The second, and the focus of this section, is angular resolution. The habitable zone of a Sun-like star lies near $1$\,au, which even for the nearest stars ($1.3$--$6$\,pc) subtends only $\sim$$0.1$--$1$\,arcsec (Table~\ref{tab:targets}). Separating a planet from its star at these separations demands both a small coronagraphic inner working angle and the highest possible angular resolution, which scales as $\lambda/D$. At $10\,\mu$m the diffraction limit is $\approx$0.3\,arcsec for a single 8.4\,m aperture, $\approx$0.4\,arcsec for the 6.5\,m JWST, but $\approx$0.1\,arcsec is currently achievable for the $22.7$\,m edge-to-edge baseline of the LBT operated in Fizeau mode. Space-based mid-infrared coronagraphy is correspondingly limited at small separations: the JWST/MIRI four-quadrant phase masks have an inner working angle of $\approx$$1\,\lambda/D$ ($\approx$0.33\,arcsec at $10.65\,\mu$m), so a $1$\,au orbit is unresolved beyond $\sim$3\,pc, restricting space-based habitable-zone imaging in the mid-infrared essentially to $\alpha$\,Centauri.\cite{boccaletti2022, rouan2000, rieke2015, gardner2006} Large ground-based apertures with high-performance coronagraphs therefore provide unique access to the habitable-zone angular scales of a broader sample of the nearest Sun-like stars. Which of those systems are most favorable has been quantified by Werber et al.\cite{werber2023}, who modeled the direct mid-infrared detectability of Earth- to giant-mass planets across the nearest FGK and BA stars --- combining each planet's predicted thermal emission with its host's luminosity, distance, and age --- and used it to prioritize the most promising habitable-zone targets; the nearest single Sun-like stars in Table~\ref{tab:targets} rank among the best.\cite{bowens2021}

\begin{table}[t]
\centering
\footnotesize
\begin{tabular*}{\textwidth}{@{\extracolsep{\fill}}lcccccccccc@{}}
\hline
 &  &  &  &  & \multicolumn{3}{c}{HZ angular radius (arcsec)} & \multicolumn{2}{c}{$a_{\rm cen}$ / element} &  \\
\cline{6-8}\cline{9-10}
Star & Sp.\ type & $d$ (pc) & $L\,(L_\odot)$ & Dec.\ ($^\circ$) & inner & center & outer & $\lambda/D$ & FWHM & LBT? \\
\hline
$\alpha$\,Cen A & G2\,V   & 1.34 & 1.52  & $-$60.8 & 0.69  & 0.92  & 1.62 & 8.4 & 11.2 & No  \\
$\alpha$\,Cen B & K1\,V   & 1.34 & 0.50  & $-$60.8 & 0.40  & 0.53  & 0.98 & 4.8 & 6.5  & No  \\
\textbf{$\epsilon$\,Eri} & K2\,V   & 3.20 & 0.32  & $-$9.5  & 0.14  & 0.18  & 0.33 & 1.6 & 2.2  & \textbf{Yes} \\
\textbf{$\tau$\,Cet}     & G8\,V   & 3.65 & 0.52  & $-$15.9 & 0.15  & 0.20  & 0.36 & 1.8 & 2.4  & \textbf{Yes} \\
$\epsilon$\,Ind & K5\,V   & 3.64 & 0.22  & $-$56.8 & 0.10  & 0.13  & 0.25 & 1.2 & 1.6  & No  \\
\textbf{61\,Cyg A }      & K5\,V   & 3.50 & 0.15  & $+$38.7 & 0.088 & 0.11  & 0.22 & 1.0 & 1.3  & \textbf{Yes} \\
\textbf{61\,Cyg B }      & K7\,V   & 3.50 & 0.085 & $+$38.7 & 0.067 & 0.083 & 0.17 & 0.8 & 1.0  & \textbf{Yes} \\
\textbf{40\,Eri A}       & K0.5\,V & 5.00 & 0.46  & $-$7.6  & 0.11  & 0.14  & 0.26 & 1.3 & 1.7  & \textbf{Yes} \\
\textbf{$\sigma$\,Dra}   & K0\,V   & 5.76 & 0.45  & $+$69.7 & 0.089 & 0.12  & 0.21 & 1.1 & 1.5  & \textbf{Yes} \\
82\,Eri         & G8\,V   & 6.04 & 0.69  & $-$43.0 & 0.11  & 0.14  & 0.25 & 1.3 & 1.7  & No  \\
\textbf{Gmb\,1618}       & K7\,V   & 4.94 & 0.16  & $+$49.8 & 0.065 & 0.081 & 0.17 & 0.7 & 1.0  & \textbf{Yes} \\
HR\,7703        & K2.5\,V & 6.00 & 0.30  & $-$36.2 & 0.072 & 0.091 & 0.18 & 0.8 & 1.1  & No  \\
\hline
\end{tabular*}
\caption{Habitable-zone (HZ) angular scales and LBT accessibility for the nearest
Sun-like stars --- the sample plotted in Fig.~\ref{fig:overview}. HZ radii are listed at three insolations, each scaled as $a\!\propto\!\sqrt{L/S_{\rm eff}}$: the Earth-equivalent (``center,'' $S_{\rm eff}\!=\!1$) distance and the optimistic inner (recent-Venus) and outer (early-Mars) limits of Kopparapu et al.~(2013).\cite{kopparapu2013, kasting1993, selsis2007, leconte2013} The center radius is also given in LBT resolution elements --- the $22.7$\,m Fizeau $\lambda/D\!\approx\!0.11$\,arcsec and the finer central-fringe FWHM ($D_{\rm eff}\!=\!30.6$\,m, $\approx\!0.082$\,arcsec). The last column marks whether the star is observable from the LBT (declination $\gtrsim\!-30^\circ$); the five southern stars ($\alpha$\,Cen\,A/B, $\epsilon$\,Ind, 82\,Eri, HR\,7703) appear as open markers in Fig.~\ref{fig:overview}. Values are approximate.}
\label{tab:targets}
\end{table}

\subsection{LBT as a pathfinder for the ELT era}
The LBT is the only operating telescope that combines two 8.4\,m apertures on a common mount, and its $22.7$\,m edge-to-edge baseline makes it the closest existing analog of a $30$\,m-class extremely large telescope. Demonstrating resolved mid-infrared habitable-zone imaging at the LBT now --- with the Q-AGPM coronagraph and the GeoSnap detector --- directly matures the techniques and hardware planned for ELT/METIS, which is expected to reach Earth-analog sensitivity around the nearest Sun-like stars in the next decade.\cite{brandl2021, bowens2021} The two upgrades described below are the critical steps toward that goal.

%%%%%%%%%%%%%%%%%%%%%%%%%%%%%%%%%%%%%%%%%%%%%%%%%%%%%%%%%%%%%%%%%%%%%%%%%%%%%%%%%
\section{THE LBTI/NOMIC INSTRUMENT}
\label{sec:nomic}

NOMIC images the $8$--$13\,\mu$m thermal infrared at the combined focus of the LBT.\cite{hoffmann2014} Ground-based thermal-infrared imaging and spectroscopy have a long heritage on 4--10\,m-class telescopes,\cite{glasse1997, kataza2000, telesco2003} which the LBT's larger collecting area and interferometric baseline extend to the angular scales of the nearest habitable zones. In its original configuration the camera uses a Raytheon $1024\times1024$ Si:As impurity-band-conduction AQUARIUS array with a $30\,\mu$m pixel pitch, giving a plate scale of $\approx$0.018\,arcsec\,pixel$^{-1}$. The AQUARIUS array, however, suffers from excess low-frequency noise (ELFN) that must be suppressed by rapid chopping, and its quantum efficiency and well depth limit the achievable signal-to-noise in the high-background ground-based regime.\cite{hoffmann2014, leisenring2023}

A key enabler for the upgrades described here is a recently commissioned LBTI observing mode in which the two 8.4\,m mirrors are used non-interferometrically as two telescopes focused onto the same detector, while the interferometric beam-directing optics are used for rapid background tracking and suppression.\cite{wagner2021spie, ertel2022} This imaging mode delivers higher sensitivity for a given exposure time and makes efficient use of both apertures. However, with ELFN demanding a rapid chopping approach, this mode could not be effectively combined with Fizeau imaging using the AQUARIUS array. A pilot survey without a coronagraph and with both sides operating independently --- the LBT Exploratory Survey for Super-Earths/Sub-Neptunes Orbiting Nearby Stars (LESSONS) --- has met its performance goals and demonstrated the feasibility,\cite{wagner2021spie,ertel2022} leaving a dedicated $N$-band coronagraph and new detector as the principal missing ingredients for reaching GNOMIC's peak performance.

%%%%%%%%%%%%%%%%%%%%%%%%%%%%%%%%%%%%%%%%%%%%%%%%%%%%%%%%%%%%%%%%%%%%%%%%%%%%%%%%%
\section{THE QUADRUPLE ANNULAR GROOVE PHASE MASK CORONAGRAPH}
\label{sec:qagpm}

\begin{figure}[p]
  \centering
  \begin{minipage}[t]{0.27\linewidth}\centering
    {\footnotesize\textbf{(a) White-light image}}\\[3pt]
    \includegraphics[width=\linewidth]{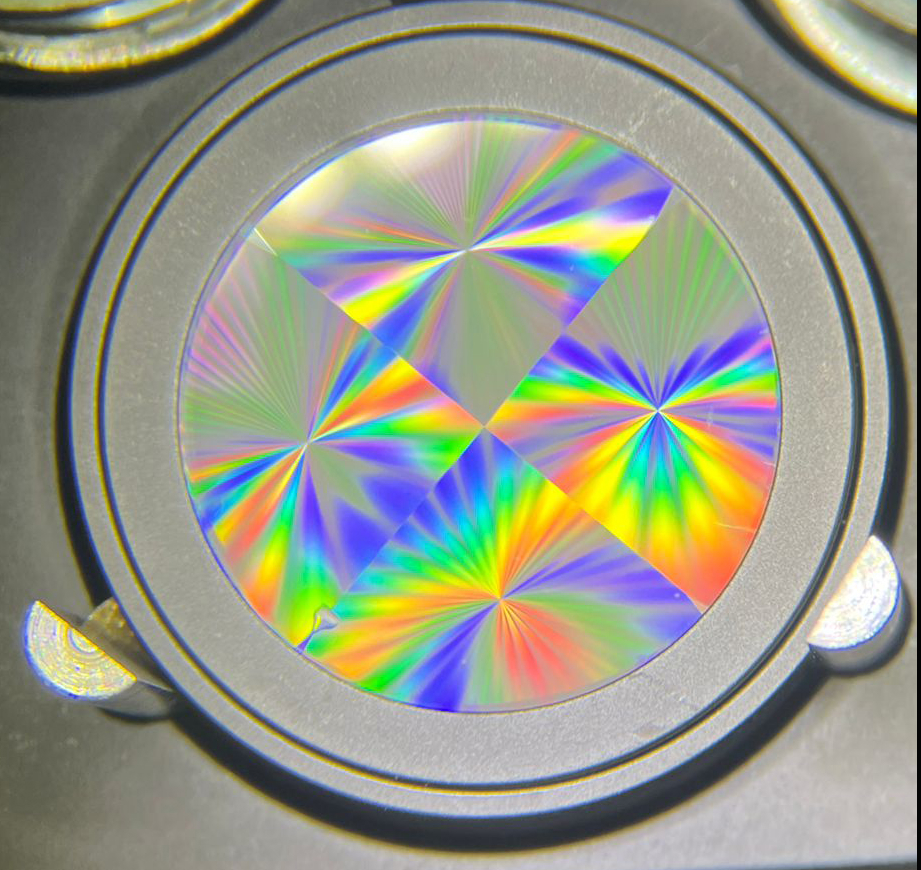}
  \end{minipage}\hspace{3pt}
  \begin{minipage}[t]{0.25\linewidth}\centering
    {\footnotesize\textbf{(b) Transmission map}}\\[3pt]
    \includegraphics[width=\linewidth]{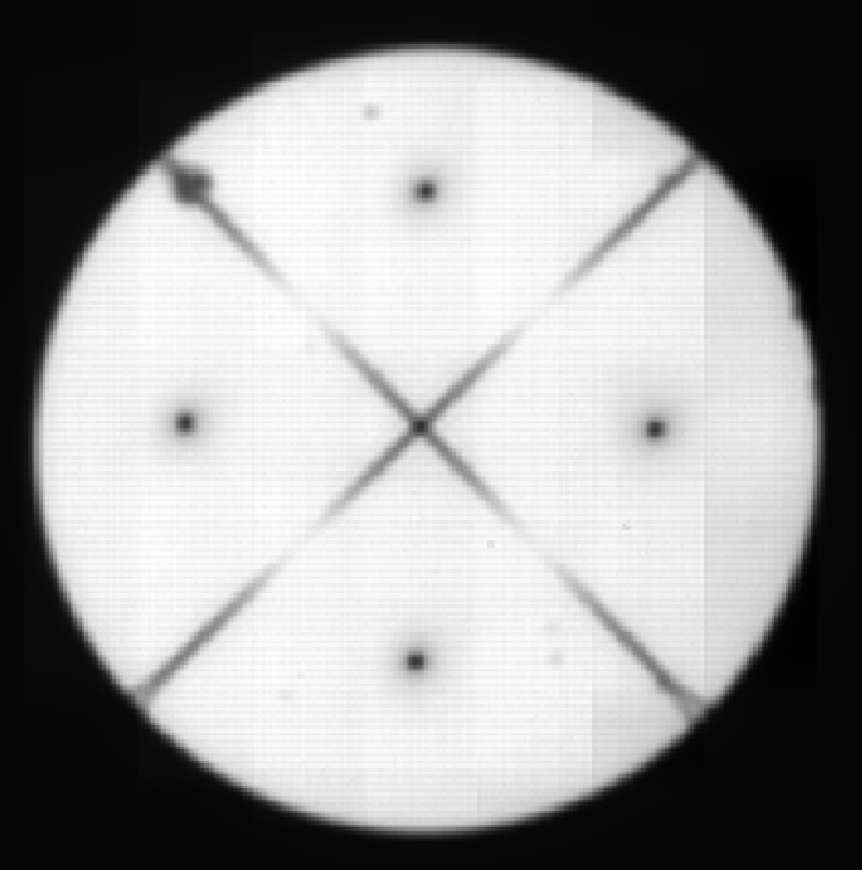}
  \end{minipage}\hfill
  \begin{minipage}[t]{0.43\linewidth}
    \caption[photo]{\label{fig:hardware}
The completed quadruple AGPM (diamond substrate) installed at the CEA Saclay test-bench. (a) Under white light the 20\,mm diamond shows vivid diffraction colors from the four subwavelength-grating AGPMs. (b) The measured transmission map reveals the four AGPM centers, the interfaces between quadrants, and the alignment mark to the upper left.}
  \end{minipage}
\end{figure}

\begin{figure}[p]
  \centering
  \begin{minipage}[t]{0.36\linewidth}\centering
    {\footnotesize\textbf{(a) Mask layout}}\\[3pt]
    \includegraphics[width=\linewidth]{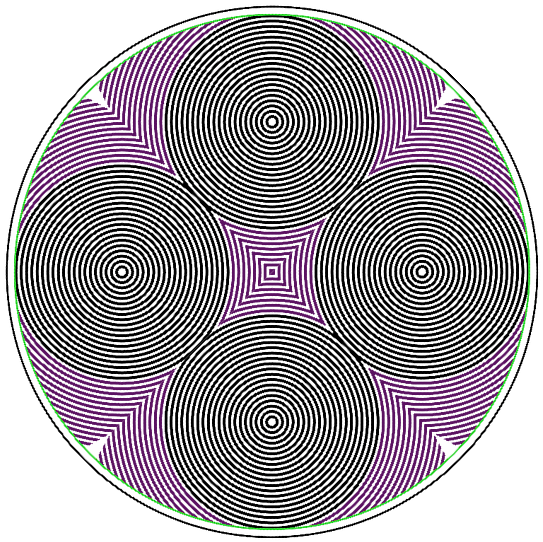}
  \end{minipage}\hfill
  \begin{minipage}[t]{0.62\linewidth}\centering
    {\footnotesize\textbf{(b) On-axis rejection spectrum}}\\[3pt]
    \includegraphics[width=\linewidth]{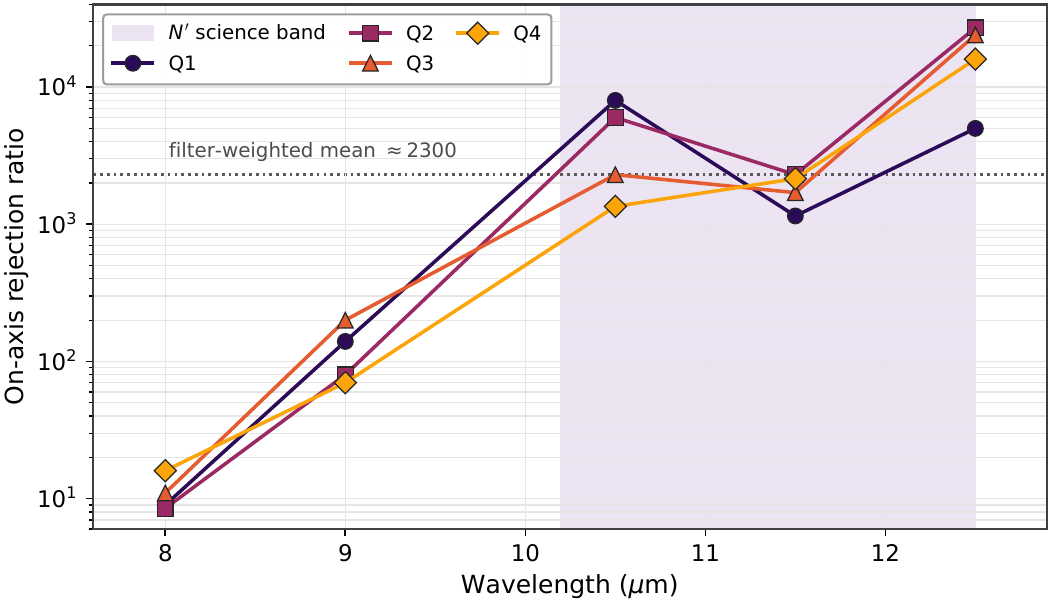}
  \end{minipage}
  \caption[Q-AGPM]
{\label{fig:qagpm} (a) The quadruple AGPM design: four annular groove phase masks etched on a single 20\,mm diamond substrate.\cite{forsberg2026} (b) Measured on-axis rejection ratio of the four quadrants versus wavelength after the depth-reduction re-etch (November 2025 CEA Saclay data); the rejection climbs from $\sim$10 below the $\approx$9.5\,$\mu$m zero-order cutoff to $10^{3}$--$10^{4}$ across the $N'$ science band (shaded), where the filter-weighted mean is $\approx$2300 (dotted). Panel (b) measured by S.~Ronayette and E.~Pantin (CEA Saclay).}
\end{figure}

\begin{figure}[p]
  \centering
  {\footnotesize\textbf{Coronagraphic Test-Bench Laser PSFs across the $N$ band}}\\[3pt]
  \includegraphics[width=\linewidth]{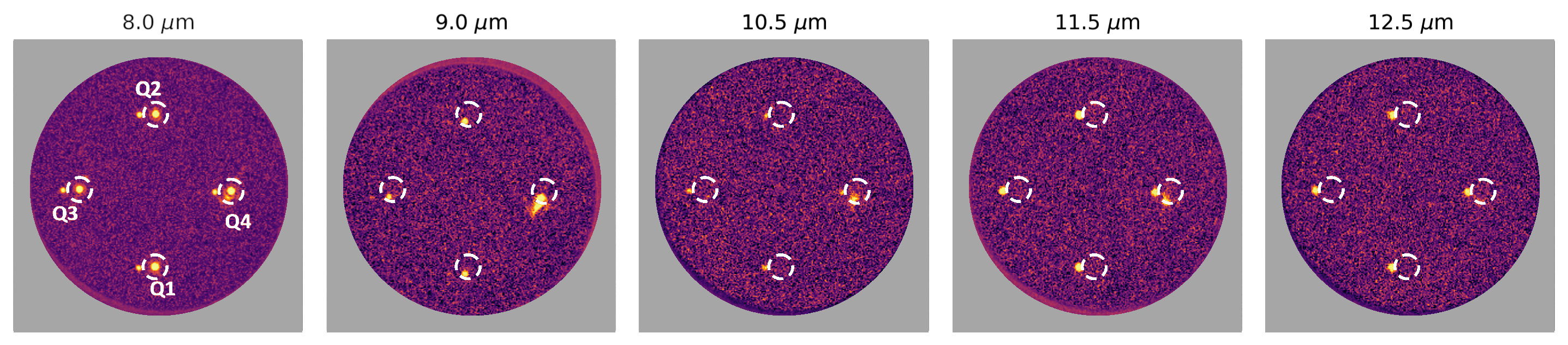}
  \caption[psf nulls]{\label{fig:psfnull}
Coronagraphic test-bench laser point-spread images of the four AGPM quadrants (Q1--Q4) at $8.0$, $9.0$, $10.5$, $11.5$, and $12.5\,\mu$m (left to right); dashed circles mark the on-axis PSF positions, fixed from the $8.0\,\mu$m panel. Below the $\approx$9.5\,$\mu$m zero-order cutoff the quadrant PSFs are only weakly suppressed; across the $N'$ band the on-axis star is extinguished, leaving only the faint cold-density ghost to the left. Measurements by S.~Ronayette and E.~Pantin (CEA Saclay).}
\end{figure}

Here, we provide an overview of the Q-AGPM and its initial performance. The detailed design, manufacturing, and lab-tested performance will be published in Forsberg et al. (in review).\cite{forsberg2026} 

\subsection{Principle and motivation}

The annular groove phase mask (AGPM) is a focal-plane vector-vortex coronagraph formed by a circular, concentric subwavelength grating etched into synthetic diamond.\cite{mawet2005, delacroix2013, foo2005} Among the many stellar coronagraph architectures --- from the classical Lyot coronagraph\cite{lyot1939} to band-limited\cite{kuchner2002} and apodized-pupil Lyot\cite{soummer2005} designs --- the vortex is distinguished by a small inner working angle and near-unity off-axis throughput. The grating behaves as a continuously rotating half-wave plate, imposing an optical vortex that diverts on-axis starlight outside the re-imaged pupil, where it is rejected by a Lyot stop, while off-axis companions are transmitted with high throughput. Diamond AGPMs have extensive on-sky and laboratory heritage --- including the $L'$-band mask offered at VLT/NACO, the $N$-band mask used for the NEAR campaign, the $L$- and $M$-band masks for LBTI,\cite{lbtiagpm} and the masks for the upcoming ELT/METIS instrument.\cite{delacroix2013, absil2013, mawet2013} The vortex's throughput and its sensitivity to low-order aberrations are well characterized,\cite{mawet2010, ruane2018} and the concept has demonstrated on-sky exoplanet imaging at small angles.\cite{mawet2010palomar, serabyn2010} The AGPM offers a small inner working angle of $\approx$$0.9\,\lambda/D$ --- the off-axis separation at which the coronagraph transmits half of a point source's flux --- with high off-axis throughput, set by a matched cold Lyot stop that rejects the diffracted on-axis starlight. 

Both NEAR and LBTI/NOMIC use AQUARIUS arrays, which are dominated by low-frequency noise. This is mitigated in the observing approach by chopping the stellar position on the detector at $\sim$1--10\,Hz and differencing frames. For a binary such as $\alpha$\,Centauri the two stars can be nodded back and forth on one mask, but for a single star, chopping on one AGPM would waste roughly half of the integration time. Placing two AGPMs adjacent to one another and chopping the star between them recovers that efficiency. Because the LBTI imaging mode uses both apertures, each of the two beams needs its own pair of masks --- four AGPMs in total. This is the motivation for the quadruple AGPM. In the Fizeau imaging mode that will be used with the GeoSnap the star is no longer chopped. For Fizeau imaging, the four quadrants instead provide redundancy, allowing the best-performing AGPM centers or detector regions to be selected, or a four-point nod to be averaged. Combining the vortex coronagraph with Fizeau recombination is a novel configuration to be validated on sky. Because the AGPM reshapes the pupil illumination downstream of the mask in a way that depends on the differing residual aberrations of the two apertures, the fringe contrast --- and hence the Fizeau gain --- may be reduced, and the realized benefit remains to be demonstrated.

A further advantage of implementing the vortex at LBTI is that the Q-AGPM sits downstream of an existing cold pupil stop. In the NEAR experiment no cold stop could be placed ahead of the mask, so thermal emission from the telescope structure was redirected into a bright halo around the vortex center, an effect referred to as the vortex center glow.\cite{shinde2022} At LBTI this glow is suppressed, adding an estimated $20$--$30\%$ in sensitivity at small separations. The approach also builds on established heritage: LBTI already operates a $3$--$5\,\mu$m double-AGPM with matched Lyot stops for warm young exoplanets,\cite{lbtiagpm} so the alignment and cold-stop procedures for the $N$-band quadruple mask are a direct extension of a proven configuration. Chopping, when used, is performed with closed-loop cryogenic pupil mirrors located after the adaptive-optics wavefront sensor, so the correction is never frozen during background measurement --- a further efficiency gain over NEAR, which has to freeze its correction for roughly half of each cycle.

\subsection{Design and fabrication}
The Q-AGPM places the four masks on a single $20$\,mm-diameter, $300\,\mu$m-thick polycrystalline diamond substrate (Figs.~\ref{fig:hardware} and~\ref{fig:qagpm}a), which guarantees that all four masks share as nearly identical a process history as possible and simplifies handling and mounting relative to four separate components.\cite{forsberg2026} Within the usable $\approx$19.4\,mm aperture, the four masks have individual diameters (and center-to-center spacing) of $\approx$8.04\,mm; outside the circular fields of view the gratings are extended to fill each quadrant and are treated as non-critical. The grating was optimized for the LBTI $N'$ filter band ($\approx$10.2--12.5\,$\mu$m) using rigorous coupled-wave analysis of the trapezoidal grooves, with a fixed period of $4\,\mu$m so that the grating remains zero-order down to $\approx$9.5\,$\mu$m (for diamond, $n=2.38$); an additional subwavelength antireflection grating was etched on the rear face, modeled to hold the backside reflectivity below $\sim$1\% across $10$--$12.7\,\mu$m. Fabrication was carried out at the {\AA}ngstr{\"o}m Laboratory, following the process developed for the METIS $N$-band masks: silicon hard-mask deposition, electron-beam lithography, solvent-assisted micromolding to transfer the pattern to diamond, and inductively coupled plasma etching, with a parallel monitor sample cross-sectioned by scanning electron microscopy to calibrate the etch depth.\cite{forsberg2026} The grating geometry was optimized following the framework established for $L$-band diamond AGPMs.\cite{vargascatalan2016}

\subsection{Cryogenic testing and performance}
We measured the Q-AGPM's null depth on a cryogenic bench at CEA Saclay, using quantum-cascade-laser sources spanning $8$--$12.5\,\mu$m. An initial round of testing, fit against the simulated null-depth spectra, indicated that the as-etched grating was slightly too deep. A controlled depth-reduction step (protecting the groove bottoms with resist and etching the exposed tops in oxygen plasma, removing $\approx$1\,$\mu$m of depth) was then applied. In the second, final round of testing (CEA Saclay, November 2025) the in-band rejection rose sharply: all four quadrants reached on-axis rejection ratios of order $10^{3}$--$10^{4}$ at $10.5$, $11.5$, and $12.5\,\mu$m (Fig.~\ref{fig:qagpm}b), in agreement with rigorous coupled-wave modeling of the grating and corresponding to a filter-weighted rejection of $\approx$2300 (null depth $\approx$$4.3\times10^{-4}$).\cite{forsberg2026} Table~\ref{tab:qagpmspec}(a) presents the Q-AGPM design and measured parameters. As expected, the rejection falls steeply below the zero-order cutoff near $9.5\,\mu$m. The same behavior is apparent in the coronagraphic test-bench laser point-spread images taken at the best-rejection positions (Fig.~\ref{fig:psfnull}): below the cutoff the four quadrant PSFs are only weakly suppressed, whereas across the $N'$ band the on-axis star is highly extinguished and only the faint cold-density ghost remains. An unexpected ghost noted on quadrant Q4 was traced to likely surface dust rather than a fabrication defect: the component has since been inspected and cleaned in a cleanroom --- four small pieces of foreign-object debris were removed from the coated face --- and is now vacuum-sealed and ready to be installed at the LBT. %On sky, the stellar point-spread function (PSF) will be held on each vortex center using post-coronagraphic tip-tilt sensing.\cite{huby2015}

%%%%%%%%%%%%%%%%%%%%%%%%%%%%%%%%%%%%%%%%%%%%%%%%%%%%%%%%%%%%%%%%%%%%%%%%%%%%%%%%%
\section{THE GEOSNAP DETECTOR UPGRADE}
\label{sec:geosnap}

The second upgrade replaces NOMIC's legacy Si:As AQUARIUS array with a $13\,\mu$m-cutoff Teledyne GeoSnap detector, a longwave HgCdTe device read out by Teledyne's GeoSnap-18 focal-plane module. A prototype GeoSnap for the MIRAC-5 instrument on the MMT --- among the first such mid-infrared arrays produced for astronomical instrumentation --- has been characterized in the laboratory and on sky and is now performing routine observations.\cite{mirac5} A GeoSnap detector will also be used in ELT/METIS\cite{brandl2021}. This successful demonstration and synergy with future ELT instrumentation motivated procurement of a similar device for LBTI.\cite{leisenring2023, cabrera2019, rogalski2005, beletic2008, rieke2007, finger2014}

\subsection{Excess low-frequency noise and the case for HgCdTe}
The principal obstacle to high-sensitivity Fizeau imaging with the AQUARIUS array is the excess low-frequency noise (ELFN) intrinsic to the Si:As detector.\cite{arrington1998} Suppressing ELFN demands rapid chopping ($\sim$10\,Hz), which is fundamentally incompatible with interferometry: maintaining fringe lock requires the stellar image to be held at a fixed position on the detector, whereas chopping repeatedly moves it. A detector free of ELFN therefore unlocks the combination of long, stable integrations and closed-loop phase control that Fizeau imaging requires. HgCdTe does not exhibit ELFN, so with the GeoSnap the star can be parked at a single location for several-minute stares, interrupted only by $\sim$30\,s telescope nods for classical background subtraction; the resulting $\sim$90\% on-source efficiency slightly exceeds the $\sim$85\% delivered by the chopped AQUARIUS system.

Laboratory characterization of the MIRAC-5 GeoSnap array shows near-ideal noise behavior out to more than a thousand co-added frames (several minutes).\cite{ertel2022, leisenring2023} A $\sim$30\% rise from residual $1/f$ noise appears after a few tens of frames, but this is far milder than the factor of $>$5 noise growth exhibited by AQUARIUS over comparable averaging. The HgCdTe quantum efficiency (QE; $\sim$0.8) also exceeds that of AQUARIUS ($\sim$0.6), but this improves sensitivity by only $\sim$15\% (combined with the $1/f$ noise).\cite{leisenring2023} The larger gains come from the observing strategy that the ELFN-free detector enables --- primarily the elimination of chopping. Taken together, these detector-enabled factors give an otherwise NEAR-like configuration a factor of $\sim$2.3 in sensitivity.

\subsection{The GeoSnap-18 focal-plane array}
The GeoSnap-18 module pairs the HgCdTe layer with a $2048\times2048$, $18\,\mu$m-pitch read-out integrated circuit (ROIC) built around a capacitive trans-impedance amplifier (CTIA) unit cell with two selectable gains. Its snapshot-shutter, integrate-while-read architecture captures flux at nearly 100\% duty cycle, and the module provides a digital interface with an on-chip 14-bit ADC, programmable windowing, and anti-blooming. Table~\ref{tab:geosnap}(b) lists the parameters most relevant to high-contrast imaging, including its dual-gain full wells and read noise, frame rates up to 85\,Hz, and 75--85\% quantum efficiency. The detector has been delivered and will undergo characterization at the Arizona InfraRed Detector Lab (AIRD Lab) over Summer 2026 before deployment to LBTI.\cite{leisenring2026}

\begin{table}[t]
\centering
\footnotesize \setlength{\tabcolsep}{4pt}
\begin{minipage}[t]{0.48\linewidth}\centering
  \textbf{(a) Q-AGPM $N'$-band coronagraph}\\[4pt]
  \begin{tabular}{ll}
  \hline
  Parameter & Value \\
  \hline
  Coronagraph type     & Vector vortex (charge 2) \\
  Substrate            & Diamond, 20\,mm\,$\times$\,300\,$\mu$m \\
  Masks / substrate    & 4 (one per quadrant) \\
  Grating period       & $4\,\mu$m (subwavelength) \\
  Zero-order cutoff    & $\approx$9.5\,$\mu$m \\
  Optimized band       & $N'$, 10.2--12.5\,$\mu$m \\
  On-axis rejection    & $10^{3}$--$10^{4}$ /quadrant \\
  Filter-weighted mean & $\approx$2300 ($4.3\times10^{-4}$) \\
  Inner working angle  & $\approx$0.9\,$\lambda/D$ \\
  Fabrication          & {\AA}ngstr{\"o}m Lab, 2025 \\
  \hline
  \end{tabular}
\end{minipage}\hfill
\begin{minipage}[t]{0.50\linewidth}\centering
  \textbf{(b) GeoSnap-18 focal-plane array}\\[4pt]
  \begin{tabular}{ll}
  \hline
  Parameter & Value \\
  \hline
  Detector material       & HgCdTe, $13\,\mu$m cutoff \\
  Array format            & $2048\times2048$ \\
  Pixel pitch             & $18\,\mu$m \\
  Unit cell / readout     & CTIA, dual-gain, 14-bit ADC \\
  Full-frame rate         & $\leq$\,85\,Hz \\
  Integration efficiency  & $>$\,97\% at 30\,Hz \\
  Quantum efficiency      & 75\,/\,85\% (min/typ.) \\
  Full well (hi/lo gain)  & $1.8\times10^{5}$\,/\,$2.6\times10^{6}\,e^-$ \\
  ROIC read noise         & 40\,/\,400\,$e^-$ (hi/lo) \\
  Operating temperature   & $\sim$45\,K \\
  \hline
  \end{tabular}
\end{minipage}
\caption{Parameters of the two principal upgrades. (a) Quadruple AGPM (Q-AGPM) $N'$-band vortex coronagraph: design parameters and cryogenic performance (fabricated at the {\AA}ngstr{\"o}m Laboratory; measured at CEA Saclay, November 2025).\cite{forsberg2026} (b) GeoSnap-18 focal-plane array parameters relevant to LBTI/NOMIC.}
\label{tab:geosnap}\label{tab:qagpmspec}
\end{table}

\subsection{Expected sensitivity and error budget}
\label{sec:budget}
Combined with Fizeau imaging, the background-limited sensitivity is expected to improve by a factor of $\gtrsim$6 relative to NEAR for a given exposure time.\cite{wagner2021spie} This gain decomposes into its multiplicative contributions: a factor of two from eliminating chopping, $1.45$ from the $2.1\times$ larger collecting area, $1.25$ from the lower thermal background at the high, cold Mount Graham site, $1.15$ from the net quantum-efficiency improvement (the raw $\sim$33\% QE gain is conservatively reduced by the residual $1/f$ noise), $\sqrt{3}$ from concentrating the Fizeau signal into a three-times-smaller area, and $\sqrt{2}$ from noise terms present in NEAR but absent at LBTI (primarily the central glow). These factors compound to a raw improvement of $\sim$$10\times$, of which we conservatively adopt $\gtrsim$$6\times$. NEAR reached an S/N\,=\,3 point-source sensitivity of $\approx$300\,$\mu$Jy in 100 hours at a separation of three resolution elements. The adopted gain, together with $\sim$100--200 hours of accumulated, multi-epoch integration, should bring the S/N\,=\,3 limit to $\approx$30--50\,$\mu$Jy, the predicted flux of a $\sim$2\,$R_\oplus$ temperate super-Earth around $\epsilon$\,Eridani. Giant planets are considerably more accessible, and are interesting targets due to prospects for detecting exomoons.\cite{wagner2025,winterhalder2026}

The dominant uncertainty in this budget is the residual $1/f$ noise, measured on the MIRAC-5 GeoSnap and conservatively estimated here at 30\% of the photon noise. The interferometric terms are well constrained: the differential tip-tilt jitter is $\approx$10\,mas RMS and the phase jitter $\approx$650\,nm ($0.37$\,rad at $11.11\,\mu$m), both measured with LBTI's PhaseCam fringe tracker and predominantly on timescales longer than the $50$--$100$\,ms NOMIC frame time, so they can be removed by re-centering individual frames.\cite{ertel2022}

\subsection{Integration and commissioning}
A feasibility study confirmed that the GeoSnap can be integrated into NOMIC with no redesign of the fore-optics. The detector package ($100\times100\times20$\,mm) closely matches that of AQUARIUS, and a solid-model mock-up shows that, mounted in place of the existing array on the NOMIC detector stage, it clears surrounding hardware by at least $0.65$\,inch on all sides while picking up existing holes on the stage plate. A machined slot and a $\sim$1\,m harness through the nearest cryostat pass-through, terminated in a potted feedthrough, route the signals out of the Nulling and Imaging Cryostat (NIC). The relaxed operating temperature ($\sim$45\,K versus $\sim$7\,K for AQUARIUS) further eases integration. Commissioning is proceeding first in the laboratory --- developing readout patterns and operations software and characterizing dark current and bias --- after which the detector will be installed in NIC during the 2026 Summer monsoon shutdown along with the Q-AGPM. An incoming inspection of the delivered device was carried out, yielding no major concerns. Installation at the LBT is planned for late Summer / early Fall 2026, with first light on sky by end of 2026.

%A few cosmetic scratches and several instances of possible foreign-object debris were noted on the detector cover, and the flex connector was found not fully seated; these items were flagged for correction before cold testing. 

\subsection{Laboratory test cryostat}
To characterize the GeoSnap and develop its operation ahead of installation, the program has procured a dedicated closed-cycle test cryostat from Infrared Laboratories (Tucson, Arizona). A Sumitomo CH-208R cold head with an F-70L water-cooled compressor cools the $\sim$$12\,$inch-diameter by $22$-inch anodized vacuum case, reaching below $30$\,K on the second-stage cold plate while holding $\pm$25\,mK stability over an hour-long test cycle. The first- and second-stage heat exchangers are oxygen-free copper and the radiation shield is gold-plated 6061 aluminum with an Aeroglaze interior and interchangeable field-of-view baffles; two motorized, eight-position cryogenic filter/aperture wheels select among $1$-inch optics. A device-under-test stage mounted to the second stage carries a copper plate with a heater and temperature sensor and a pedestal that mates to the GeoSnap ``Spider'' package, and the vacuum cover seals a $50.8$\,mm window. The dewar lets us measure quantum efficiency, $1/f$ and read noise, dark current, and operating temperature, and to exercise the readout electronics, before the detector is integrated into NIC. After integration, further lab development is enabled by a bare MUX. 

%%%%%%%%%%%%%%%%%%%%%%%%%%%%%%%%%%%%%%%%%%%%%%%%%%%%%%%%%%%%%%%%%%%%%%%%%%%%%%%%%
\section{FIZEAU IMAGING AND HIGH-CONTRAST RESULTS}
\label{sec:fizeau}

\begin{figure}[t]
  \centering
  \includegraphics[width=\linewidth]{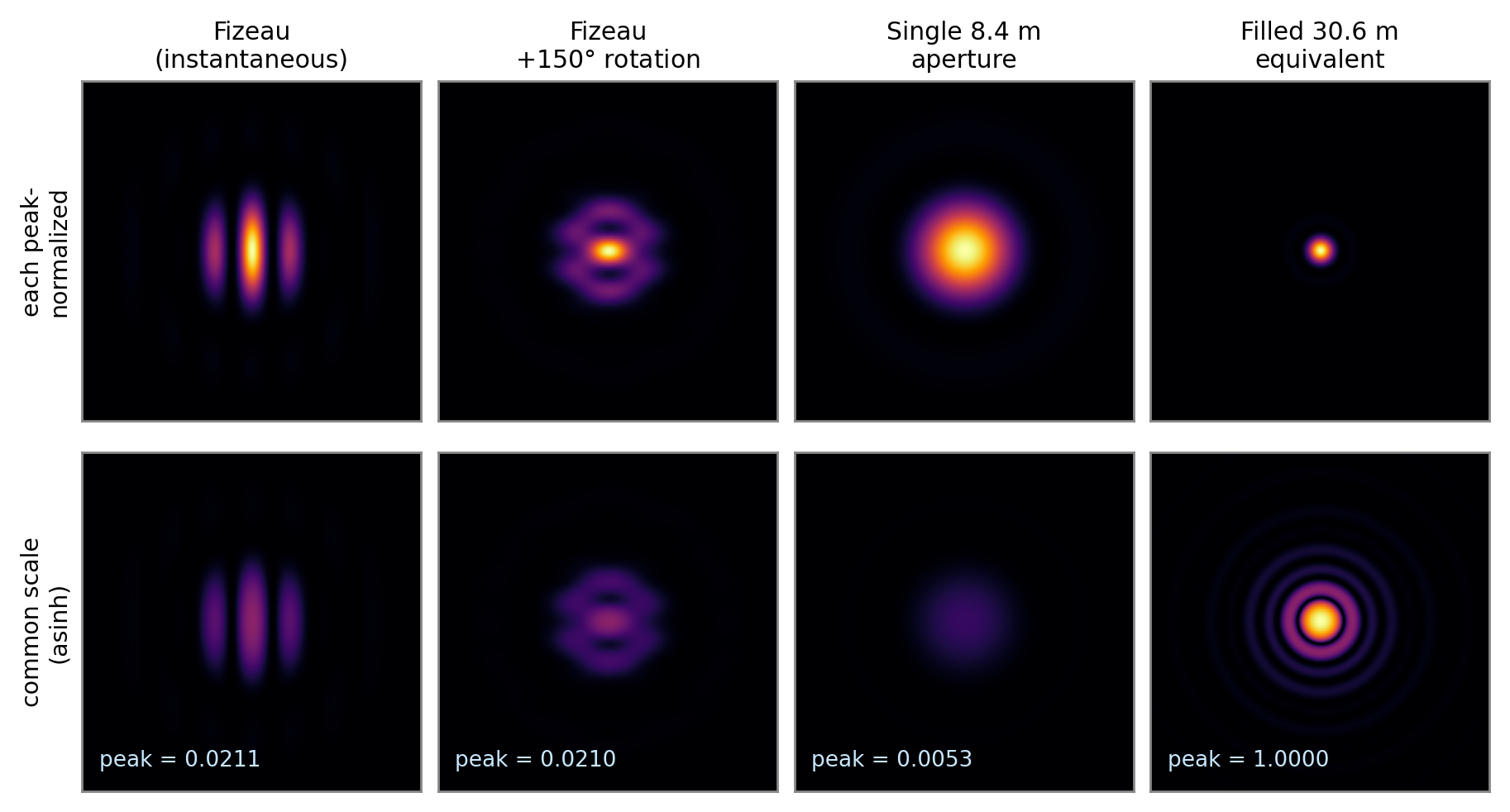}
  \caption[fizeau psf]{\label{fig:fizeaupsf}
The LBTI Fizeau point-spread function at $N$ band ($11\,\mu$m), computed for the working pupil. \emph{Columns, left to right:} the instantaneous fringed PSF (fringe period $\lambda/B\approx0.16$\,arcsec); the same PSF synthesized over $150^{\circ}$ of field rotation; a single 8.4\,m aperture; and a filled $\sim$30\,m aperture matching the Fizeau central-fringe width. \emph{Top row:} each panel normalized to its own peak (shape comparison). \emph{Bottom row:} the same four on a common photometric scale (same star and exposure, asinh stretch), with peak intensities annotated --- co-phasing and rotation synthesis sharpen the core, while the filled-aperture equivalent (peak $\equiv$1) still concentrates the most light.}

  \vspace{0.9em}
  \includegraphics[width=\linewidth]{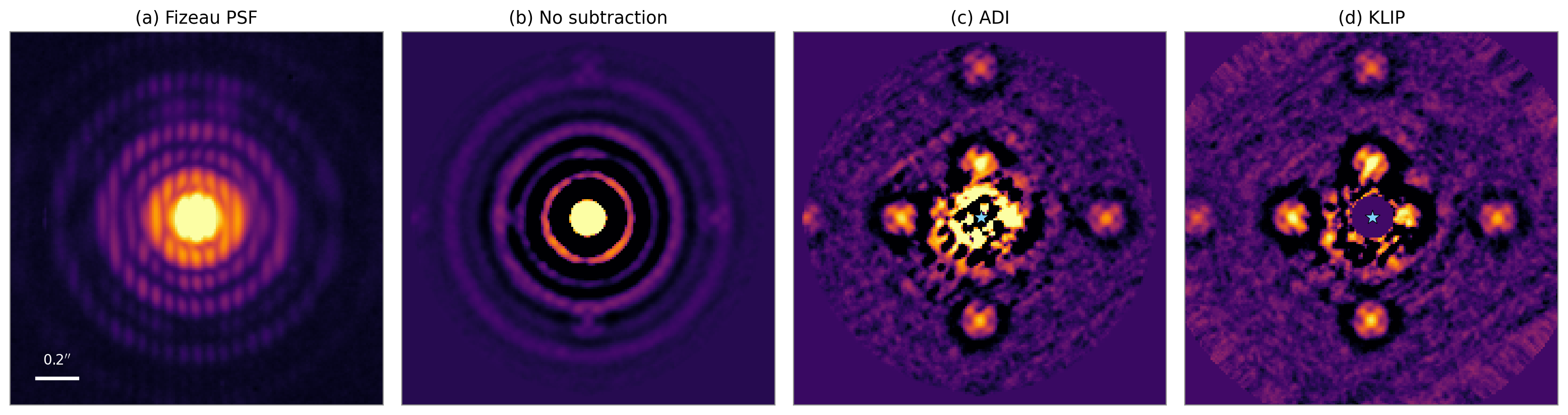}
  \caption[progression]{\label{fig:progression}
Reduced images from the demonstration high-contrast Fizeau sequence obtained with the new FFTCam fringe tracker (a $12.3$-minute LBTI/LMIRCam $M$-band, $4.77\,\mu$m dataset spanning $115^{\circ}$ of field rotation), with seven point sources injected at separations of $0.15$--$0.87$\,arcsec (in steps of $0.12$\,arcsec) and contrasts, respectively, of $1.0$, $0.70$, $0.50$, $0.38$, $0.27$, $0.20$, and $0.14\times10^{-3}$. \emph{(a)} The real fringed Fizeau PSF (cf.\ Fig.~\ref{fig:fizeaupsf}). \emph{(b)} The unsubtracted, derotated stack, in which the inner injected planets are buried under the stellar halo and its concentric diffraction rings. \emph{(c)} Classical angular differential imaging\cite{marois2006} and \emph{(d)} KLIP-ADI\cite{soummer2012} subtraction, which recovers the innermost injected source at 0.15 arcsec.}
\end{figure}

Both hardware upgrades are ultimately in service of high-contrast imaging in the LBTI Fizeau mode, whose point-spread function sets the achievable resolution and contrast. In Fizeau mode the two 8.4\,m apertures are co-phased across the 14.4\,m center-to-center baseline, so the combined point-spread function is the single-aperture envelope multiplied by interference fringes with a period of $\lambda/B\approx0.16$\,arcsec at $11\,\mu$m (Fig.~\ref{fig:fizeaupsf}). The maximum spatial frequency is set by the 22.7\,m edge-to-edge span, giving a central fringe as narrow as that of a $\sim$30\,m filled aperture --- roughly three times finer than a single 8.4\,m mirror. As the sky rotates relative to the pupil over an observation, co-adding the frames synthesizes a nearly filled-aperture core that concentrates the planet light while diluting the stellar speckle field. See also Isbell et al., in these proceedings.\cite{isbelllive}

Co-phasing the two apertures improves point-source sensitivity in three distinct ways. First, in the background-limited regime it concentrates the planet's flux into a smaller PSF core, raising its signal against the sky background. Second, in the contrast-limited regime it concentrates that flux against the stellar speckle field, which is itself suppressed in the dark valleys of the fringe pattern; because a planet samples many favorable fringe positions over an observation, the predicted contrast gain ranges from $\sim$2--10$\times$ for exposures much longer than the atmospheric coherence time to $\sim$10--100$\times$ for shorter ones, and is of order $10\times$ for NOMIC's $\sim$50\,ms $N$-band coherence time.\cite{wagner2021spie, ertel2022} Third, the dark fringe valleys extend the effective inner working angle, enabling detections inside the $\lambda/D$ of a single 8.4\,m aperture, at separations of $\sim$0.15--0.3\,arcsec --- the habitable-zone scales of $\epsilon$\,Eridani and $\tau$\,Ceti. This interferometric gain has now been demonstrated on sky. Using FFTCam,\cite{stone2026} a new fringe-tracking camera that replaces PhaseCam, we obtained a demonstration high-contrast Fizeau sequence with LBTI: a $12.3$-minute (on-source integration) $M$-band ($4.77\,\mu$m) angular-differential-imaging dataset on a bright ($m\approx4.3$ at $M$ band) star spanning $115^{\circ}$ of field rotation. %That these targets are $\sim$17$\times$ fainter than $\alpha$\,Centauri\,A further reduces the impact of stellar speckles relative to NEAR. These detector and interferometric gains compound with the post-processing applied to NOMIC data --- angular differential imaging\cite{marois2006} and principal-component-analysis subtraction of the thermal background, the latter improving contrast by a factor of $2$--$3$ down to the diffraction limit.\cite{rousseau2024}

\begin{figure}[t]
  \centering
  \includegraphics[width=\linewidth]{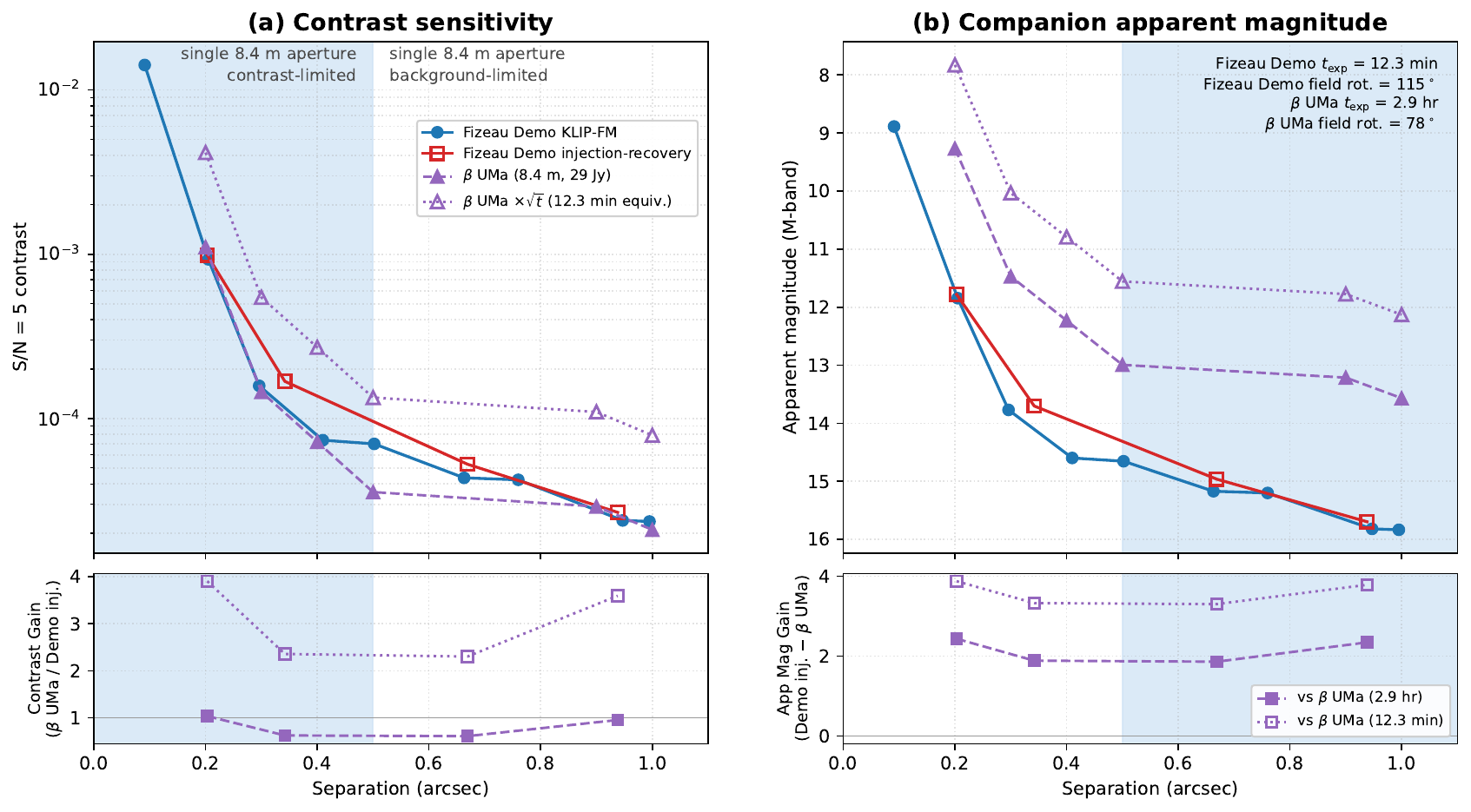}
  \caption[hpboo]{\label{fig:hpboo}
Sensitivity of the $12.3$-minute $M$-band ($4.77\,\mu$m) Fizeau demonstration (a $3.2$\,Jy, $m\approx4.3$ target), compared with the single 8.4\,m aperture $\beta$\,UMa. \emph{Top:} the as-measured S/N\,=\,5 contrast (a) and the corresponding companion apparent magnitude (b) versus separation, for the KLIP forward-model and independent injection--recovery, with $\beta$\,UMa observed for $2.9$\,hr and scaled to the same exposure time. Light-blue shading highlights the contrast-limited regime ($<0.5''$) in panel~(a) and the background-limited regime ($>0.5''$) in panel~(b). \emph{Bottom:} the resulting gain over the single-aperture observation. Contrasts are the as-measured S/N\,=\,5 sensitivities; the S/N\,=\,3 values quoted in the text are $\approx$0.6$\times$ lower.}
\end{figure}

 Figure~\ref{fig:progression} shows the reduction sequence --- the fringed Fizeau PSF, the unsubtracted derotated stack, and the cADI and KLIP results --- in which a spiral of injected point sources is cleanly recovered. After KLIP, the reduction reaches an S/N\,=\,3 contrast of $\sim$$6\times10^{-4}$ at $0.2$\,arcsec, $\sim$$6\times10^{-5}$ near $0.35$\,arcsec, and $\sim$$1\times10^{-5}$ beyond $0.9$\,arcsec (Fig.~\ref{fig:hpboo}). Relative to a single 8.4\,m aperture observed for the same time, the co-phased Fizeau reduction improves the contrast by a factor of $\sim$2--4 across this range --- an empirical confirmation of the interferometric gain at the separations relevant to habitable-zone exoplanet imaging. Although this demonstration used LMIRCam at $M$-band, the same co-phasing and rotation synthesis apply directly to the $N$-band GNOMIC observations planned with the Q-AGPM and GeoSnap. Two effects should make the $N$-band gains stronger still. First, fringe stability improves at longer wavelengths --- i.e., optical-path-difference errors are a fixed physical length, whereas the residual phase error scales as a fraction of the wavelength. This especially enhances the gain in the contrast-limited regions close to the star. Second, the background-limited regime is reached at smaller separations (in units of $\lambda/D$), so the larger background-limited gain takes over closer to the star, though not in absolute angle given the coarser resolution at $N$ band. Finally, the reductions shown here use classical ADI and KLIP; dedicated PSF-subtraction and planet-detection algorithms\cite{lafreniere2007, mugnier2009, flasseur2018, samland2021} may offer additional gains at the smallest separations.

%%%%%%%%%%%%%%%%%%%%%%%%%%%%%%%%%%%%%%%%%%%%%%%%%%%%%%%%%%%%%%%%%%%%%%%%%%%%%%%%%
\section{ROLE WITHIN THE BREAKTHROUGH WATCH PROGRAM}
\label{sec:breakthrough}

LBTI/GNOMIC is a central facility of the Breakthrough Watch program at the University of Arizona, a jointly sponsored effort with Breakthrough Initiatives, the University of Arizona, and the LBT Observatory to carry out the deepest thermal-infrared search yet for habitable-zone planets around the nearest single Sun-like stars. Where NEAR reached sub-Neptune sensitivity around the $\alpha$\,Centauri binary --- a system whose habitable zones can stably host planets\cite{quarles2016} below current radial-velocity detection limits\cite{zhao2018} --- the larger collecting area and resolution of the LBT, combined with the Q-AGPM and the GeoSnap detector, are designed to extend the approach to single stars and to lower planet masses.\cite{wagner2021, wagner2021spie, ertel2022} The program aims to complete a first census for super-Earth-sized planets and for ice/gas giants that could host rocky moons within their habitable zones.\cite{wagner2021spie, ertel2022} Planet-occurrence statistics from radial-velocity and transit surveys indicate that small and giant planets are common around Sun-like and lower-mass stars,\cite{fischer2005, petigura2013, dressing2015, fernandes2019, bryson2021} so a volume-limited search of the nearest systems is likely to yield detections. For habitable-zone imaging the ultimate astrophysical noise floor is set by the level of exozodiacal dust. LBTI is uniquely positioned here, having already measured the exozodi levels of the nearest stars through the HOSTS nulling survey, and deeming our key targets to be feasible for deep exoplanet imaging.\cite{ertel2020, wyatt2008, millangabet2011}

%\subsection{Science goals and target selection}
%The program targets the nearest single Sun-like stars where the $\sim$1\,au habitable zone is resolved at $N$ band. Candidates are selected by distance ($d\!\lesssim\!4$\,pc, so that the habitable zone subtends more than a resolution element), luminosity near solar (placing the $\sim$300\,K equilibrium temperature near $1$\,au), the presence of known or suspected planetary systems, and visibility from Mount Graham for at least half a night; younger systems are favored because their planets are warmer and brighter, although old $\sim$300\,K planets remain detectable. The resulting shortlist is led by $\epsilon$\,Eridani and $\tau$\,Ceti --- the only two with known planets --- followed by 61\,Cygni, Procyon, and a handful of others. Adopting measured planet-occurrence rates ($\sim$10--40\% of stars host a $0.3$--$13\,M_{\rm Jup}$ planet between $1$ and $5$\,au), the full campaign is expected to yield one to a few new planets, with each star carrying a $\sim$20\% chance of a super-Earth or sub-Neptune within reach. The primary quantitative goal is $\gtrsim$100 hours on $\epsilon$\,Eridani, sufficient to probe planets as small as $\sim$2\,$R_\oplus$ in its habitable zone.\cite{wagner2021spie, werber2023}

\subsection{Primary and secondary targets}
$\epsilon$\,Eridani (K2\,V, $3.2$\,pc) is among the primary targets. It hosts a known radial-velocity giant of $\sim$$0.7\,M_{\rm Jup}$ near $3.5$\,au\cite{hatzes2000} whose predicted $N$-band brightness of $\sim$0.3\,mJy makes it detectable within a few hours. Recovering it directly would break the mass--luminosity degeneracy and provide early commissioning science.\cite{werber2023} The system may host additional, as-yet-undetected planets in or near the habitable zone, where super-Earths are predicted at $\sim$30--50\,$\mu$Jy --- within reach of $\sim$100--200-hour integrations. $\tau$\,Ceti (G8\,V, $3.65$\,pc), the closest single solar-type (G) star and among the oldest ($\sim$8--10\,Gyr), presents a habitable zone spanning $\approx$0.15--0.36\,arcsec. Its radial-velocity planet candidates remain unconfirmed\cite{tuomi2013, feng2017}, but astrometry indicates a possible wide, few-Jupiter-mass companion at $3$--$20$\,au\cite{kervella2019} that would be directly imageable with LBTI. Its predicted habitable-zone super-Earths ($\sim$20--30\,$\mu$Jy) are somewhat fainter than those of $\epsilon$\,Eridani.\cite{werber2023} 61\,Cygni (K5\,V, $3.5$\,pc) has a habitable zone roughly twice as close in, at the edge of the LBT Fizeau diffraction limit. As a far-northern system it cannot be observed from the southern hemisphere, leaving LBTI uniquely able to image its habitable zone among current and near-term facilities. Procyon and additional bright nearby stars round out the sample as commissioning and mass-sensitivity targets.

\subsection{Pilot survey and multi-epoch strategy}
\begin{figure}[t]
  \centering
  \begin{minipage}[b]{0.385\linewidth}\centering
    \begin{overpic}[width=\linewidth]{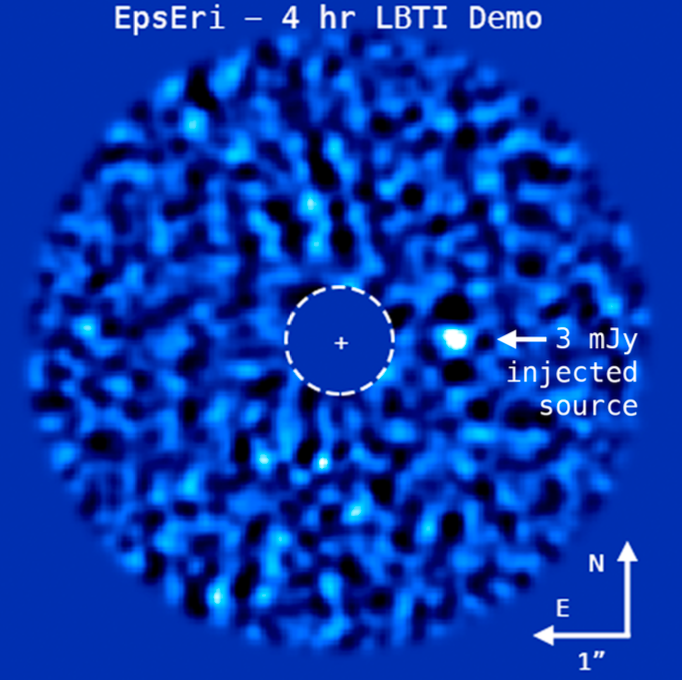}
      \put(4,90){\footnotesize\color{white}\textbf{(a)}}
    \end{overpic}
  \end{minipage}\hfill
  \begin{minipage}[b]{0.58\linewidth}\centering
    \begin{overpic}[width=\linewidth]{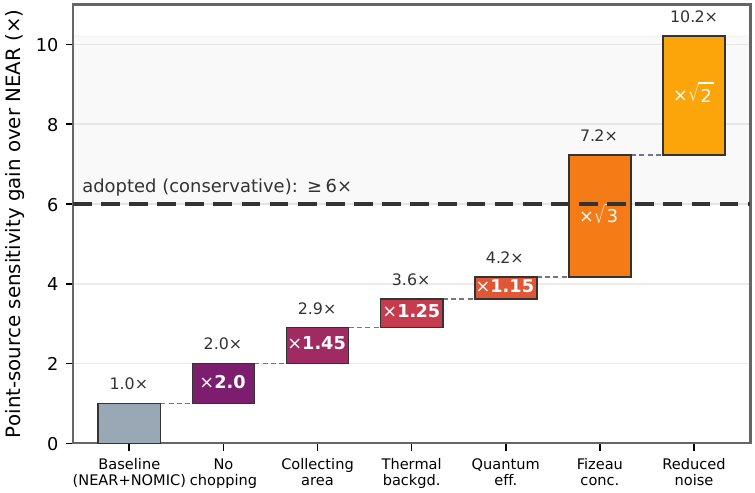}
      \put(11.5,61.5){\footnotesize\textbf{(b)}}
    \end{overpic}
  \end{minipage}
  \caption[LESSONS]{\label{fig:lessons}
(a) A 4-hour LBTI/NOMIC $N$-band ($10\,\mu$m) demonstration on $\epsilon$\,Eridani from the LESSONS pilot survey (single aperture, no coronagraph): a $3$\,mJy point source at $1''$ is recovered at S/N\,=\,3; the bar marks $1$\,arcsec. (b) Point-source sensitivity budget for GNOMIC relative to NEAR: the multiplicative factors compound to a raw $\sim$$10\times$ gain, of which we conservatively adopt $\gtrsim$$6\times$ (Sec.~\ref{sec:budget}).\cite{wagner2021spie, werber2023}}
\end{figure}

The LESSONS pilot survey, carried out from 2020--2023, has already demonstrated the approach without a coronagraph across a sample including several of the same stars ($\epsilon$\,Eridani, $\tau$\,Ceti, Vega, $\beta$\,Leonis, and others). Operating a single aperture without a coronagraph, a $4$-hour integration reaches an S/N\,=\,3 point-source sensitivity of $\approx$3\,mJy at $1''$ (Fig.~\ref{fig:lessons}a), or $\approx$0.6\,mJy scaled to $\sim$100 hours. Adding the Q-AGPM is expected to bring a single LBT aperture to roughly NEAR's single-aperture depth, so the Fizeau and detector gains of Fig.~\ref{fig:lessons}b and the resulting $\sim$30--50\,$\mu$Jy habitable-zone estimate (Sec.~\ref{sec:budget}) follow whether the projection is anchored to NEAR or to this LBTI demonstration. With Fizeau imaging, the GeoSnap detector, and $\sim$100-hour integrations, the program therefore projects sensitivity to rocky planets in the habitable zones of $\epsilon$\,Eridani and other nearby debris-hosting stars.\cite{wagner2021spie, werber2023, greaves1998} Multi-epoch data from the campaign can be combined with the K-Stacker algorithm, which recenters images along trial Keplerian orbits to recover faint orbiting planets below the single-epoch detection limit; K-Stacker has already been applied to the 100-hour NEAR data set of $\alpha$\,Centauri\cite{lecoroller2022} and is being adapted to the case of LBTI Fizeau imaging.

\subsection{Observing strategy and program status}
The upgrade is supported by a guaranteed allocation of roughly 45 nights across the coming Fall observing semesters, scheduled in a queue that gives the program the highest priority when conditions are suitable. Deep $N$-band imaging demands excellent mid-infrared conditions --- low precipitable water vapor, good seeing, and cold temperatures --- which an analysis of several years of Mount Graham weather indicates are met roughly $15\%$ of the time. Short-term water-vapor forecasting is used to trigger the most demanding observations during the driest nights (below $1$\,mm of precipitable water vapor), which are the most sensitive. Accumulating $\gtrsim$100 hours on a single target therefore requires many partial nights spread over several campaigns, which the multi-year allocation is designed to deliver. Following the model established by NEAR, the data will be made public immediately through the LBT Observatory archive, and the community is invited to work with the data.

\subsection{Complementarity with contemporaneous facilities}
For the nearest stars the LBTI Fizeau mode reaches stellocentric separations several times smaller than JWST at comparable wavelengths, making it the most sensitive facility for close-in, habitable-zone separations around these targets until the next generation of $30$\,m-class telescopes comes online.\cite{boccaletti2022} ELT/METIS is not expected to reach full science operations until the early 2030s, and it and the LBT are hemispherically complementary --- LBTI can reach far-Northern targets such as 61\,Cygni that METIS cannot, while METIS can reach $\alpha$\,Centauri and other targets in the South.\cite{brandl2021, bowens2021} The GeoSnap upgrade also broadens the instrument's reach beyond exoplanets: it improves LBTI's nulling sensitivity to exozodiacal dust by several times --- enough to help retire the exozodi risk to future space missions designed to image exo-Earths\cite{kennedy2013} --- and opens efficient $L$/$M$-band nulling as well as extragalactic and solar-system programs for which the legacy AQUARIUS array was ill-suited.\cite{ertel2020}

%%%%%%%%%%%%%%%%%%%%%%%%%%%%%%%%%%%%%%%%%%%%%%%%%%%%%%%%%%%%%%%%%%%%%%%%%%%%%%%%%
\section{SUMMARY AND STATUS}
\label{sec:summary}

We have described two complementary upgrades to LBTI/NOMIC. The quadruple annular groove phase mask provides the camera's first dedicated $N$-band coronagraph; by placing four AGPMs on a single diamond substrate it matches the LBTI dual-aperture imaging mode and supports efficient chopping for a single star. The mask has been fabricated at the {\AA}ngstr{\"o}m Laboratory and characterized at CEA Saclay --- reaching per-quadrant rejection ratios of $10^{3}$--$10^{4}$ across the $N'$ band after a depth-reduction re-etch (mean $\approx$2300). The $13\,\mu$m-cutoff GeoSnap detector replaces the legacy AQUARIUS array with a higher-QE, larger-well-depth sensor that is crucially free of excess low-frequency noise, enabling sensitive Fizeau imaging. A high-contrast demonstration Fizeau imaging sequence, obtained on sky with LBTI's new FFTCam fringe tracker, reaches an S/N\,=\,3 contrast of $\sim$$1\times10^{-5}$, demonstrating a factor of $\sim$2--4 contrast gain over an equal-time single aperture. Both upgrades are scheduled for installation at the LBT in Summer/Fall 2026, with first light on sky by the end of 2026. On-sky commissioning results and the Breakthrough Watch campaign observations will immediately follow. Together these upgrades and the capabilities they enable are expected to deliver substantially deeper contrast at the small angular separations corresponding to the habitable zones of the nearest single Sun-like stars, advancing the Breakthrough Watch goal of imaging temperate planets in our immediate solar neighborhood.

%%%%%%%%%%%%%%%%%%%%%%%%%%%%%%%%%%%%%%%%%%%%%%%%%%%%%%%%%%%%%%%%%%%%%%%%%%%%%%%%%
\acknowledgments
This material is based upon work supported by the U.S.\ National Science Foundation under award No. 2406668. The GeoSnap detector upgrade and the associated survey are carried out as part of the Breakthrough Watch program sponsored by the Breakthrough Initiatives. This material is based upon work supported by the National Aeronautics and Space Administration under agreement No. 80NSSC21K0593 for the program ``Alien Earths.'' O.A.\ is a Research Director of the Fonds de la Recherche Scientifique -- FNRS. The nanostructure mastering is partially funded by the Research Council of Finland Flagship Programme, Photonics Research and Innovation (PREIN), decision number 346518. The LBT is an international collaboration among institutions in the United States and Europe. At the time data were acquired for this research, LBT Corporation Members were the University of Arizona on behalf of the Arizona Board of Regents; Istituto Nazionale di Astrofisica, Italy; and The Ohio State University, representing The Ohio State University, University of Notre Dame, University of Minnesota, and University of Virginia. This research used the facilities of the Italian Center for Astronomical Archives (IA2) operated by INAF at the Astronomical Observatory of Trieste. Observations have benefited from the use of ALTA Center (alta.arcetri.inaf.it) forecasts performed with the Astro-Meso-Nh model. Initialization data of the ALTA automatic forecast system come from the General Circulation Model (HRES) of the European Centre for Medium Range Weather Forecasts. We acknowledge the use of the Large Binocular Telescope Interferometer (LBTI) and the support from the LBTI team, specifically from Jon Reese and Jared Carlson.

%The program also includes a winter workshop on mid-infrared coronagraphy and exoplanet imaging, hosted at the University of Arizona's Biosphere~2.
% TODO before submission: add grant numbers, the Breakthrough Foundation
% acknowledgment, Q-AGPM fabrication and CEA Saclay collaborators, and any
% facility/funder statements required by SPIE.

%%%%%%%%%%%%%%%%%%%%%%%%%%%%%%%%%%%%%%%%%%%%%%%%%%%%%%%%%%%%%%%%%%%%%%%%%%%%%%%%%
% References (inline \bibitem; no BibTeX run needed). references.bib mirrors these.
%%%%%%%%%%%%%%%%%%%%%%%%%%%%%%%%%%%%%%%%%%%%%%%%%%%%%%%%%%%%%%%%%%%%%%%%%%%%%%%%%


\begin{thebibliography}{99}

\bibitem{astro2020}
National Academies of Sciences, Engineering, and Medicine, {\it Pathways to Discovery in Astronomy and Astrophysics for the 2020s}, The National Academies Press, Washington, DC (2021).

\bibitem{exostrategy2018}
National Academies of Sciences, Engineering, and Medicine, {\it Exoplanet Science Strategy}, The National Academies Press, Washington, DC (2018).

\bibitem{chauvin2004}
G.~Chauvin et al., ``A giant planet candidate near a young brown dwarf: direct VLT/NACO observations using IR wavefront sensing,'' {\it Astron.\ Astrophys.} {\bf 425}, L29 (2004).

\bibitem{marois2008}
C.~Marois et al., ``Direct imaging of multiple planets orbiting the star HR 8799,'' {\it Science} {\bf 322}, 1348 (2008).

\bibitem{macintosh2015}
B.~Macintosh et al., ``Discovery and spectroscopy of the young jovian planet 51 Eri b with the Gemini Planet Imager,'' {\it Science} {\bf 350}, 64 (2015).

\bibitem{haffert2019}
S.~Y. Haffert et al., ``Two accreting protoplanets around the young star PDS 70,'' {\it Nat.\ Astron.} {\bf 3}, 749 (2019).

\bibitem{quanz2015}
S.~P. Quanz et al., ``Direct detection of exoplanets in the 3--10\,$\mu$m range with E-ELT/METIS,'' {\it Int.\ J.\ Astrobiol.} {\bf 14}(2), 279--289 (2015).

\bibitem{kasper2017}
M.~Kasper et al., ``NEAR: Low-mass planets in $\alpha$\,Cen with VISIR,'' {\it The Messenger} {\bf 169}, 16--20 (2017).

\bibitem{wagner2021}
K.~Wagner, A.~Boehle, P.~Pathak, et al., ``Imaging low-mass planets within the habitable zone of $\alpha$\,Centauri,'' {\it Nat.\ Commun.} {\bf 12}, 922 (2021).

\bibitem{brandl2021}
B.~Brandl et al., ``METIS: the mid-infrared ELT imager and spectrograph,'' {\it The Messenger} {\bf 182}, 22--26 (2021).

\bibitem{quanz2022}
S.~P. Quanz et al., ``Large Interferometer For Exoplanets (LIFE). I. Improved exoplanet detection yield estimates for a large mid-infrared space-interferometer mission,'' {\it Astron.\ Astrophys.} {\bf 664}, A21 (2022).

\bibitem{hoffmann2014}
W.~F. Hoffmann et al., ``Operation and performance of the mid-infrared camera, NOMIC, on the Large Binocular Telescope,'' {\it Proc.\ SPIE} {\bf 9147}, 91471O (2014).

\bibitem{hinz2016}
P.~M. Hinz et al., ``Overview of LBTI: a multipurpose facility for high spatial resolution observations,'' {\it Proc.\ SPIE} {\bf 9907}, 990704 (2016).

\bibitem{ertel2020spie}
S.~Ertel, P.~M. Hinz, J.~M. Stone, et al., ``Overview and prospects of the LBTI beyond the completed HOSTS survey,'' {\it Proc.\ SPIE} {\bf 11446}, 1144607 (2020).

\bibitem{ertel2020}
S.~Ertel et al., ``The HOSTS survey for exozodiacal dust: observational results from the complete survey,'' {\it Astron.\ J.} {\bf 159}, 177 (2020).

\bibitem{spalding}
E.~Spalding, P.~M. Hinz, S.~Ertel, et al., ``Towards controlled Fizeau observations with the Large Binocular Telescope,'' {\it Proc.\ SPIE} {\bf 10701}, 107010J (2018).

\bibitem{isbell2025}
J.~W. Isbell, S.~Ertel, et al., ``Direct imaging of active galactic nucleus outflows and their origin with the 23\,m Large Binocular Telescope,'' {\it Nat.\ Astron.} {\bf 9}, 417 (2025).

\bibitem{labeyrie1996}
A.~Labeyrie, ``Resolved imaging of extra-solar planets with future 10--100\,km optical interferometric arrays,'' {\it Astron.\ Astrophys.\ Suppl.\ Ser.} {\bf 118}, 517 (1996).

\bibitem{monnier2003}
J.~D. Monnier, ``Optical interferometry in astronomy,'' {\it Rep.\ Prog.\ Phys.} {\bf 66}, 789 (2003).

\bibitem{ertel2022}
S.~Ertel et al., ``Imaging nearby, habitable-zone planets with the Large Binocular Telescope Interferometer,'' {\it Proc.\ SPIE} {\bf 12183}, 1218302 (2022).

\bibitem{wagner2021spie}
K.~Wagner, S.~Ertel, J.~Stone, J.~Leisenring, D.~Apai, M.~Kasper, O.~Absil, L.~Close, D.~Defr\`ere, O.~Guyon, and J.~Males, ``Imaging low-mass planets within the habitable zones of nearby stars with ground-based mid-infrared imaging,'' {\it Proc.\ SPIE} {\bf 11823}, 118230G (2021).

\bibitem{forsberg2026}
P.~Forsberg, S.~Ronayette, K.~Wagner, S.~Ertel, M.~Fetisova, P.~Karvinen, M.~Kuittinen, E.~Pantin, O.~Absil, and M.~Karlsson, ``The quad annular groove phase mask coronagraph for the Large Binocular Telescope Interferometer,'' {\it J.\ Astron.\ Telesc.\ Instrum.\ Syst.} (in review, 2026).

\bibitem{macintosh2014}
B.~Macintosh et al., ``First light of the Gemini Planet Imager,'' {\it Proc.\ Natl.\ Acad.\ Sci.\ U.S.A.} {\bf 111}, 12661 (2014).

\bibitem{hinkley2011}
S.~Hinkley et al., ``A new high-contrast imaging program at Palomar Observatory,'' {\it Publ.\ Astron.\ Soc.\ Pac.} {\bf 123}, 74 (2011).

\bibitem{desmarais2002}
D.~J. Des Marais et al., ``Remote sensing of planetary properties and biosignatures on extrasolar terrestrial planets,'' {\it Astrobiology} {\bf 2}, 153 (2002).

\bibitem{kaltenegger2017}
L.~Kaltenegger, ``How to characterize habitable worlds and signs of life,'' {\it Annu.\ Rev.\ Astron.\ Astrophys.} {\bf 55}, 433 (2017).

\bibitem{schwieterman2018}
E.~W. Schwieterman et al., ``Exoplanet biosignatures: a review of remotely detectable signs of life,'' {\it Astrobiology} {\bf 18}, 663 (2018).

\bibitem{cash2006}
W.~Cash, ``Detection of Earth-like planets around nearby stars using a petal-shaped occulter,'' {\it Nature} {\bf 442}, 51 (2006).

\bibitem{mamajek2024}
E.~E. Mamajek and K.~R. Stapelfeldt, ``NASA Exoplanet Exploration Program (ExEP) mission star list for the Habitable Worlds Observatory (2023),'' {\it arXiv:2402.12414} (2024).

\bibitem{kopparapu2013}
R.~K. Kopparapu et al., ``Habitable zones around main-sequence stars: new estimates,'' {\it Astrophys.\ J.} {\bf 765}, 131 (2013).

\bibitem{bracewell1978}
R.~N. Bracewell, ``Detecting nonsolar planets by spinning infrared interferometer,'' {\it Nature} {\bf 274}, 780 (1978).

\bibitem{leger1996}
A.~L\'eger et al., ``Could we search for primitive life on extrasolar planets in the near future? The DARWIN project,'' {\it Icarus} {\bf 123}, 249 (1996).

\bibitem{boccaletti2022}
A.~Boccaletti et al., ``JWST/MIRI coronagraphic performances as measured on-sky,'' {\it Astron.\ Astrophys.} {\bf 667}, A165 (2022).

\bibitem{rouan2000}
D.~Rouan et al., ``The four-quadrant phase-mask coronagraph. I. Principle,'' {\it Publ.\ Astron.\ Soc.\ Pac.} {\bf 112}, 1479 (2000).

\bibitem{rieke2015}
G.~H. Rieke et al., ``The Mid-Infrared Instrument for the James Webb Space Telescope, I: Introduction,'' {\it Publ.\ Astron.\ Soc.\ Pac.} {\bf 127}, 584 (2015).

\bibitem{gardner2006}
J.~P. Gardner et al., ``The James Webb Space Telescope,'' {\it Space Sci.\ Rev.} {\bf 123}, 485 (2006).

\bibitem{werber2023}
Z.~Werber, K.~Wagner, and D.~Apai, ``The direct mid-infrared detectability of habitable-zone exoplanets around nearby stars,'' {\it Astron.\ J.} {\bf 165}, 133 (2023).

\bibitem{bowens2021}
R.~Bowens et al., ``Exoplanets with ELT-METIS. I. Estimating the direct imaging exoplanet yield around stars within 6.5 parsecs,'' {\it Astron.\ Astrophys.} {\bf 653}, A8 (2021).

\bibitem{kasting1993}
J.~F. Kasting, D.~P. Whitmire, and R.~T. Reynolds, ``Habitable zones around main sequence stars,'' {\it Icarus} {\bf 101}, 108 (1993).

\bibitem{selsis2007}
F.~Selsis et al., ``Habitable planets around the star Gliese 581?,'' {\it Astron.\ Astrophys.} {\bf 476}, 1373 (2007).

\bibitem{leconte2013}
J.~Leconte et al., ``Increased insolation threshold for runaway greenhouse processes on Earth-like planets,'' {\it Nature} {\bf 504}, 268 (2013).

\bibitem{glasse1997}
A.~C.~H. Glasse, E.~Ettedgui-Atad, and J.~W. Harris, ``Michelle mid-infrared spectrometer and imager,'' {\it Proc.\ SPIE} {\bf 2871}, 1197 (1997).

\bibitem{kataza2000}
H.~Kataza et al., ``COMICS: the cooled mid-infrared camera and spectrometer for the Subaru telescope,'' {\it Proc.\ SPIE} {\bf 4008}, 1144 (2000).

\bibitem{telesco2003}
C.~M. Telesco et al., ``CanariCam: a multimode mid-infrared camera for the Gran Telescopio Canarias,'' {\it Proc.\ SPIE} {\bf 4841}, 913 (2003).

\bibitem{leisenring2023}
J.~M. Leisenring et al., ``Evaluating the GeoSnap $13\,\mu$m cutoff HgCdTe detector for mid-IR ground-based astronomy,'' {\it Astron.\ Nachr.} {\bf 344}, e20230103 (2023).

\bibitem{mawet2005}
D.~Mawet, P.~Riaud, O.~Absil, and J.~Surdej, ``Annular groove phase mask coronagraph,'' {\it Astrophys.\ J.} {\bf 633}, 1191--1200 (2005).

\bibitem{delacroix2013}
C.~Delacroix et al., ``Laboratory demonstration of a mid-infrared AGPM vector vortex coronagraph,'' {\it Astron.\ Astrophys.} {\bf 553}, A98 (2013).

\bibitem{foo2005}
G.~Foo, D.~M. Palacios, and G.~A. Swartzlander, ``Optical vortex coronagraph,'' {\it Opt.\ Lett.} {\bf 30}, 3308 (2005).

\bibitem{lyot1939}
B.~Lyot, ``The study of the solar corona and prominences without eclipses,'' {\it Mon.\ Not.\ R.\ Astron.\ Soc.} {\bf 99}, 580 (1939).

\bibitem{kuchner2002}
M.~J. Kuchner and W.~A. Traub, ``A coronagraph with a band-limited mask for finding terrestrial planets,'' {\it Astrophys.\ J.} {\bf 570}, 900 (2002).

\bibitem{soummer2005}
R.~Soummer, ``Apodized pupil Lyot coronagraphs for arbitrary telescope apertures,'' {\it Astrophys.\ J.} {\bf 618}, L161 (2005).

\bibitem{lbtiagpm}
D.~Defr\`ere, O.~Absil, P.~M. Hinz, et al., ``L$'$-band AGPM vector vortex coronagraph's first light on LBTI/LMIRCam,'' {\it Proc.\ SPIE} {\bf 9148}, 91483X (2014).

\bibitem{absil2013}
O.~Absil et al., ``Searching for companions down to 2 AU from $\beta$ Pictoris using the $L'$-band AGPM coronagraph on VLT/NACO,'' {\it Astron.\ Astrophys.} {\bf 559}, L12 (2013).

\bibitem{mawet2013}
D.~Mawet et al., ``Ring-apodized vortex coronagraphs for obscured telescopes. I. Transmissive ring apodizers,'' {\it Astrophys.\ J.\ Suppl.} {\bf 209}, 7 (2013).

\bibitem{mawet2010}
D.~Mawet et al., ``The vector vortex coronagraph: sensitivity to central obscuration, low-order aberrations, chromaticism, and polarization,'' {\it Proc.\ SPIE} {\bf 7739}, 773914 (2010).

\bibitem{ruane2018}
G.~Ruane et al., ``Review of high-contrast imaging systems for current and future ground- and space-based telescopes. I. Coronagraph design methods and optical performance metrics,'' {\it Proc.\ SPIE} {\bf 10698}, 106982S (2018).

\bibitem{mawet2010palomar}
D.~Mawet et al., ``The vector vortex coronagraph: laboratory results and first light at Palomar Observatory,'' {\it Astrophys.\ J.} {\bf 709}, 53 (2010).

\bibitem{serabyn2010}
E.~Serabyn, D.~Mawet, and R.~Burruss, ``An image of an exoplanet separated by two diffraction beamwidths from a star,'' {\it Nature} {\bf 464}, 1018 (2010).

\bibitem{shinde2022}
M.~Shinde, C.~Delacroix, G.~Orban~de~Xivry, O.~Absil, and R.~van~Boekel, ``Modeling the vortex center glow in the ELT/METIS vortex coronagraph,'' {\it Proc.\ SPIE} {\bf 12187}, 121870E (2022).

\bibitem{vargascatalan2016}
E.~Vargas Catal\'an et al., ``Optimizing the subwavelength grating of $L$-band annular groove phase masks for high coronagraphic performance,'' {\it Astron.\ Astrophys.} {\bf 595}, A127 (2016).

\bibitem{mirac5}
R.~Bowens, J.~M. Leisenring, M.~R. Meyer, et al., ``Commissioning of the MIRAC-5 mid-infrared instrument on the MMT,'' {\it Publ.\ Astron.\ Soc.\ Pac.} {\bf 137}(1), 014401 (2025).

\bibitem{cabrera2019}
M.~S. Cabrera et al., ``Development of $13\,\mu$m cutoff HgCdTe detector arrays for astronomy,'' {\it J.\ Astron.\ Telesc.\ Instrum.\ Syst.} {\bf 5}(3), 036005 (2019).

\bibitem{rogalski2005}
A.~Rogalski, ``HgCdTe infrared detector material: history, status and outlook,'' {\it Rep.\ Prog.\ Phys.} {\bf 68}, 2267 (2005).

\bibitem{beletic2008}
J.~W. Beletic et al., ``Teledyne Imaging Sensors: infrared imaging technologies for astronomy and civil space,'' {\it Proc.\ SPIE} {\bf 7021}, 70210H (2008).

\bibitem{rieke2007}
G.~H. Rieke, ``Infrared detector arrays for astronomy,'' {\it Annu.\ Rev.\ Astron.\ Astrophys.} {\bf 45}, 77 (2007).

\bibitem{finger2014}
G.~Finger et al., ``SAPHIRA detector for infrared wavefront sensing,'' {\it Proc.\ SPIE} {\bf 9148}, 914817 (2014).

\bibitem{arrington1998}
D.~C. Arrington, J.~E. Hubbs, M.~E. Gramer, and G.~A. Dole, ``Impact of excess low-frequency noise (ELFN) in Si:As impurity band conduction (IBC) focal plane arrays for astronomical applications,'' {\it Proc.\ SPIE} {\bf 3379}, 361 (1998).

\bibitem{leisenring2026}
J.~M. Leisenring et al., ``A GeoSnap HgCdTe detector upgrade for LBTI/NOMIC,'' {\it Proc.\ SPIE}, in press (2026).

\bibitem{wagner2025}
K.~Wagner, E.~S. Douglas, S.~Ertel, et al., ``Astrometric methods for detecting exomoons orbiting imaged exoplanets: prospects for detecting moons orbiting a giant planet in $\alpha$\,Centauri A's habitable zone,'' {\it Astrophys.\ J.\ Lett.} {\bf 991}, L44 (2025).

\bibitem{winterhalder2026}
T.~O. Winterhalder, A.~M\'erand, J.~Kammerer, et al., ``Astrometric exomoon detection by means of optical interferometry,'' {\it Astron.\ Astrophys.} {\bf 705}, A216 (2026).

\bibitem{marois2006}
C.~Marois, D.~Lafreni\`ere, R.~Doyon, B.~Macintosh, and D.~Nadeau, ``Angular differential imaging: a powerful high-contrast imaging technique,'' {\it Astrophys.\ J.} {\bf 641}, 556--564 (2006).

\bibitem{soummer2012}
R.~Soummer, L.~Pueyo, and J.~Larkin, ``Detection and characterization of exoplanets and disks using projections on Karhunen-Lo\`eve eigenimages,'' {\it Astrophys.\ J.\ Lett.} {\bf 755}(2), L28 (2012).

\bibitem{isbelllive}
J.~W. Isbell, S.~Ertel, F.~Pedichini, et al., ``LBT Interferometer Visible Extension (LIVE): pioneering thirty-meter-class telescope imaging in the visible,'' {\it Proc.\ SPIE} {\bf 14148}, in press (2026).

\bibitem{stone2026}
J.~M. Stone et al., ``FFTCam: a Fizeau fringe-tracking camera for the LBTI,'' {\it Proc.\ SPIE}, in press (2026).

\bibitem{lafreniere2007}
D.~Lafreni\`ere et al., ``A new algorithm for point-spread function subtraction in high-contrast imaging: a demonstration with angular differential imaging,'' {\it Astrophys.\ J.} {\bf 660}, 770 (2007).

\bibitem{mugnier2009}
L.~M. Mugnier et al., ``Optimal method for exoplanet detection by angular differential imaging,'' {\it J.\ Opt.\ Soc.\ Am.\ A} {\bf 26}, 1326 (2009).

\bibitem{flasseur2018}
O.~Flasseur et al., ``Exoplanet detection in angular differential imaging by statistical learning of the nonstationary patch covariances: the PACO algorithm,'' {\it Astron.\ Astrophys.} {\bf 618}, A138 (2018).

\bibitem{samland2021}
M.~Samland et al., ``TRAP: a temporal systematics model for improved direct detection of exoplanets at small angular separations,'' {\it Astron.\ Astrophys.} {\bf 646}, A24 (2021).

\bibitem{quarles2016}
B.~Quarles and J.~J. Lissauer, ``Long-term stability of planets in the $\alpha$\,Centauri system,'' {\it Astron.\ J.} {\bf 151}, 111 (2016).

\bibitem{zhao2018}
L.~Zhao et al., ``Planet detectability in the $\alpha$\,Centauri system,'' {\it Astron.\ J.} {\bf 155}, 24 (2018).

\bibitem{fischer2005}
D.~A. Fischer and J.~Valenti, ``The planet-metallicity correlation,'' {\it Astrophys.\ J.} {\bf 622}, 1102 (2005).

\bibitem{petigura2013}
E.~A. Petigura, A.~W. Howard, and G.~W. Marcy, ``Prevalence of Earth-size planets orbiting Sun-like stars,'' {\it Proc.\ Natl.\ Acad.\ Sci.\ U.S.A.} {\bf 110}, 19273 (2013).

\bibitem{dressing2015}
C.~D. Dressing and D.~Charbonneau, ``The occurrence of potentially habitable planets orbiting M dwarfs estimated from the full Kepler dataset and an empirical measurement of the detection sensitivity,'' {\it Astrophys.\ J.} {\bf 807}, 45 (2015).

\bibitem{fernandes2019}
R.~B. Fernandes et al., ``Hints for a turnover at the snow line in the giant planet occurrence rate,'' {\it Astrophys.\ J.} {\bf 874}, 81 (2019).

\bibitem{bryson2021}
S.~Bryson et al., ``The occurrence of rocky habitable-zone planets around solar-like stars from Kepler data,'' {\it Astron.\ J.} {\bf 161}, 36 (2021).

\bibitem{wyatt2008}
M.~C. Wyatt, ``Evolution of debris disks,'' {\it Annu.\ Rev.\ Astron.\ Astrophys.} {\bf 46}, 339 (2008).

\bibitem{millangabet2011}
R.~Millan-Gabet et al., ``Exozodiacal dust levels for nearby main-sequence stars: a survey with the Keck Interferometer Nuller,'' {\it Astrophys.\ J.} {\bf 734}, 67 (2011).

\bibitem{hatzes2000}
A.~P. Hatzes et al., ``Evidence for a long-period planet orbiting $\epsilon$\,Eridani,'' {\it Astrophys.\ J.} {\bf 544}, L145 (2000).

\bibitem{tuomi2013}
M.~Tuomi et al., ``Signals embedded in the radial velocity noise: periodic variations in the $\tau$\,Ceti velocities,'' {\it Astron.\ Astrophys.} {\bf 551}, A79 (2013).

\bibitem{feng2017}
F.~Feng et al., ``Color difference makes a difference: four planet candidates around $\tau$\,Ceti,'' {\it Astron.\ J.} {\bf 154}, 135 (2017).

\bibitem{kervella2019}
P.~Kervella, F.~Arenou, F.~Mignard, and F.~Th\'evenin, ``Stellar and substellar companions of nearby stars from Gaia DR2: binarity from proper motion anomaly,'' {\it Astron.\ Astrophys.} {\bf 623}, A72 (2019).

\bibitem{greaves1998}
J.~S. Greaves et al., ``A dust ring around $\epsilon$\,Eridani: analog to the young solar system,'' {\it Astrophys.\ J.} {\bf 506}, L133 (1998).

\bibitem{lecoroller2022}
H.~Le Coroller et al., ``Efficiently combining $\alpha$\,Cen\,A multi-epoch high-contrast imaging data: application of K-Stacker to the 80-hour NEAR campaign,'' {\it Astron.\ Astrophys.} {\bf 667}, A142 (2022).

\bibitem{kennedy2013}
G.~M. Kennedy and M.~C. Wyatt, ``The bright end of the exo-zodi luminosity function: disc evolution and implications for exo-Earth detectability,'' {\it Mon.\ Not.\ R.\ Astron.\ Soc.} {\bf 433}, 2334 (2013).

% --- references below are not currently cited (citing text commented out); uncomment to restore ---
%\bibitem{maire2020}
%A.-L. Maire et al., ``Design, pointing control, and on-sky performance of the mid-infrared vortex coronagraph for the VLT/NEAR experiment,'' {\it J.\ Astron.\ Telesc.\ Instrum.\ Syst.} {\bf 6}, 035003 (2020).

%\bibitem{huby2015}
%E.~Huby et al., ``Post-coronagraphic tip-tilt sensing for vortex phase masks: the QACITS technique,'' {\it Astron.\ Astrophys.} {\bf 584}, A74 (2015).

%\bibitem{rousseau2024}
%H.~Rousseau et al., ``Improving mid-infrared thermal background subtraction with principal component analysis,'' {\it Astron.\ Astrophys.} {\bf 687}, A147 (2024).
\end{thebibliography}
\end{document}